\documentclass[
superscriptaddress,
amsmath,amssymb,
aps,
prl,
twocolumn
]{revtex4-1}

\usepackage{wrapfig}

\usepackage{tikz}
\usepackage{graphicx}
\graphicspath{{figures/}}
\usepackage{dcolumn}
\usepackage{bm}
\usepackage{hyperref}
\usepackage{multirow}
\usepackage{array}
\usepackage{booktabs}
\usepackage{ctable}
\usepackage{upgreek}
\usepackage{epsfig,psfrag,subfigure,amsopn}
\usepackage{mathrsfs}
\usepackage{amssymb}
\usepackage{amsbsy}
\usepackage{color}
\usepackage{cancel}

\usepackage{pifont}
\usepackage{marginnote}
\usepackage{float}

\definecolor{dgreen}{rgb}{0,0.7,0}

\newcommand{\sectionprl}[1]{{\em #1}\/.---}

\newcommand{\titlename}{Universal distribution of the number of minima for random walks and L\'evy flights}
\newcommand{\bea}{\begin{eqnarray}}
\newcommand{\eea}{\end{eqnarray}}

\begin{document}
	
	\title{\titlename}

%

\author{Anupam Kundu}
\address{International Centre for Theoretical Sciences, Tata Institute of Fundamental Research, Bengaluru -- 560089, India}
\author{Satya N. Majumdar}
\address{LPTMS, CNRS, Univ. Paris-Sud, Universit\'e Paris-Saclay, 91405 Orsay, France}
\author{Gr\'egory Schehr}
\address{Sorbonne Universit\'e, Laboratoire de Physique Th\'eorique et Hautes Energies, CNRS UMR 7589, 4 Place Jussieu, 75252 Paris Cedex 05, France}	
	
\date{\today}
	
\begin{abstract}
We compute exactly the full distribution of the number $m$ of local minima in a one-dimensional landscape generated by a random walk or a L\'evy flight. We consider two different ensembles of landscapes, one with a fixed number of steps $N$ and the other till the first-passage time of the random walk to the origin. We show that  the distribution of $m$ is drastically different in the two ensembles (Gaussian in the former case,  while having a power-law tail $m^{-3/2}$ in the latter case). However, the most striking aspect of our results is that, in each case, the distribution is completely universal for all $m$ (and not just for large $m$), i.e., independent of the jump distribution in the random walk. This means that the distributions are exactly identical for L\'evy flights and random walks with finite jump variance. Our analytical results are in excellent agreement with our numerical simulations.  
\end{abstract}	
	
\maketitle

\sectionprl{Introduction} Estimating the number of stationary points of a random manifold is a problem of fundamental importance across fields such as physics, chemistry, mathematics and computer science~\cite{Adler, Azais, freund1995saddles, halperin1966impurity, broderix2000energy, longuet1960reflection,halperin_82,Ros_Fyo_23}. 
Counting of such stationary points appears in many contexts, such as in liquids where they represent the maxima, minima and the 
saddles of the random potential energy landscape \cite{broderix2000energy}. In glassy systems, the number of such stationary points provides a measure of the complexity (entropy) of metastable states \cite{bray1980metastable,annibale2003supersymmetric,aspelmeier2004complexity,Satya_Martin,Hivert,Sollich}. In string theory one is often interested in estimating the number of local extrema in the moduli space that represent different possible vacua \cite{aazami2006cosmology,susskind2003anthropic}. In this context, random matrix theory is often used as an important tool to count such stationary points \cite{fyodorov2004complexity,fyodorov2012critical,cavagna2000index,Dean_Satya_06,BrayDean_07,index_09,Auffinger_13,FyoKho_21}. 
The  local maxima of the phenotypic fitness landscape representing optimal phenotypes play an important role in evolutionary biology~\cite{barton2005fitness,Krug_review,Krug_fitness1,Krug_fitness2}. In optics, the estimation of the number of specular points on a random reflecting surface 
and also the electric field intensity 
of speckle laser patterns require the knowledge of the number of stationary points of a Gaussian random field \cite{longuet1960reflection,halperin_82}. Another recent application concerns data science where such stationary points play an important role in the non-convex optimization of large-dimensional data \cite{Ganguli_14}. Most studies typically focus on the mean number of stationary points of a random landscape, using for instance the Kac-Rice formula~\cite{Rice}. However, computing the full distribution of the number of stationary points, and also the number of maxima, minima and saddles, remains a formidably challenging problem, except for uncorrelated random fields for which the distribution is Gaussian by the Central Limit Theorem \cite{Satya_Martin}.  

\begin{figure}[t]
\includegraphics[width = \linewidth]{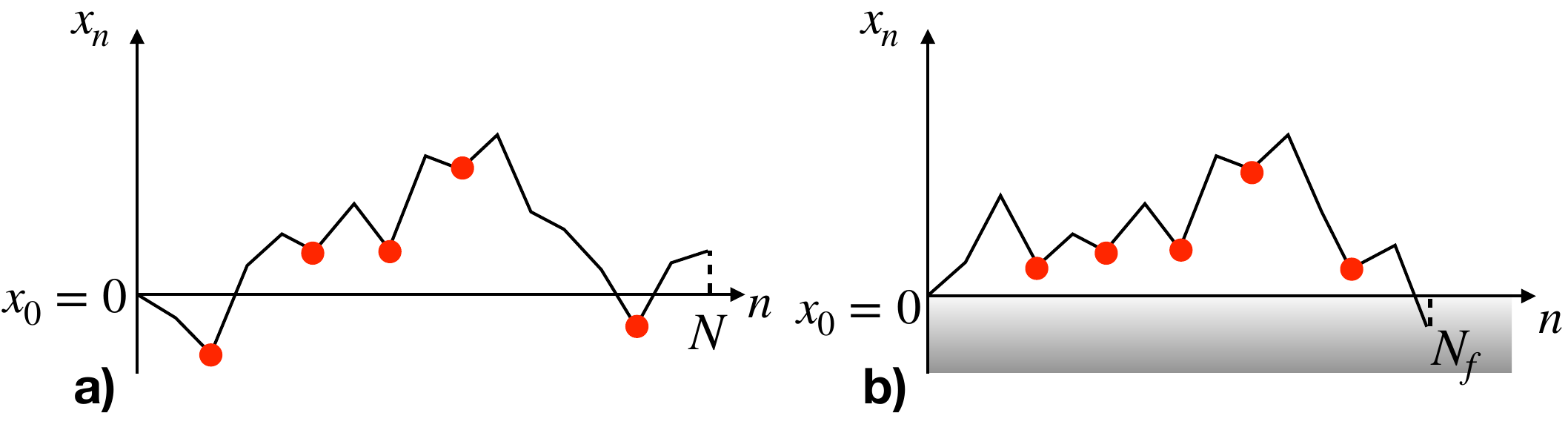}
\caption{{\bf a)}: A typical trajectory of a random walk evolving via Eq. (\ref{def_RW}) up to $N$ steps, starting at $x_0=0$. The solid red dots denote the local minima up to step $N$. {\bf b):} A typical trajectory of the same random walk as in Eq. (\ref{def_RW}) but up to step $N_f$ where it crosses its initial value from above for the first time, with the red dots indicating the local minima in the trajectory.}\label{Fig_intro}
\end{figure}

In this Letter, we consider a one-dimensional non-Gaussian landscape generated by the trajectory of a one-dimensional random walk of $N$ steps (see Fig. \ref{Fig_intro} a)). We consider a discrete-time random walker on a line, starting at the origin. Its position $x_n$ at step $n$ evolves via
\bea \label{def_RW}
x_n = x_{n-1} + \eta_n \;, \; n \geq 1 \:,
\eea
where $\eta_n$'s are independent and identically distributed random jumps, each drawn from a symmetric and continuous distribution $\phi(\eta)$. This random walk model includes L\'evy flights where $\phi(\eta) \sim 1/|\eta|^{1+\mu}$ has a power law tail with L\'evy exponent $0<\mu<2$.

This simple random walk landscape  model plays an important role in many contexts. 
This includes the celebrated Sinai's model of the transport of a single particle in a random walk landscape \cite{Sinai,BG,Comtet1,Comtet2,Monthus,Comtet3,Satya_funct,Dean}, with applications to understanding slow
dynamics in glassy disordered systems as well as in biology \cite{Nelson1,Nelson2,Oshanin}. The number of maxima of such a random walk landscape is precisely the number of barriers that the particle has to cross and the statistics of this number plays an important role in the slow dynamics of the particle.  
Similarly, the total number of local minima corresponds to the number of troughs where the particle can get trapped. 

Another well known system is $1+1$-dimensional discrete solid-on-solid (SOS) models, defined on a lattice of size $N$, that are known to converge in their stationary state to precisely a random walk trajectory of $N$ steps given by Eq. (\ref{def_RW}). In this case the effective noise distribution is directly related to the nearest neighbour interaction in the SOS model \cite{us_SOS}. Computing the statistics of the number of local maxima and minima for such stationary interfaces are important to characterize the roughness of surface fluctuations, with interesting applications in massively parallel algorithms for discrete-event simulations~\cite{Zia_00}. In this context, the mean number of such stationary points for different discrete interface models 
has been computed~\cite{Zia_00}, but its full distribution still remains elusive. 

Yet another application is the trajectory of a continuous-time run and tumble particle (RTP) in $d$-dimensions where a particle like E-coli bacteria, starting from the origin, chooses a velocity ${\bf v}$ drawn from an arbitrary isotropic distribution $W(|{\bf v}|)$ and moves ballistically during a random time \textcolor{black}{drawn from an exponential distribution (with rate $\gamma$) and then tumbles instantaneously (i.e., it chooses a new velocity from $W(|{\bf v}|)$).} The runs and tumblings alternate \cite{Berg,TC2008,ReviewRTP_1,ReviewRTP_2,Mori,Mori_PRE}. The $x$-component of this $d$-dimensional continuous-time process can be mapped onto a discrete-time random walk of $N$ steps, where the number of tumblings $N-1$ is a random variable, given the duration $t$ \cite{Mori,Mori_PRE}.  It is natural to ask how many of these tumblings in time $t$ result in a direction reversal of the particle. This is precisely the number of stationary points of the underlying random walk landscape.

In addition to the statistics of the number of maxima/minima of such a random walk landscape of fixed $N$ steps, it is also interesting to study these questions for a random walk till its first-passage time to its starting point, see Fig. \ref{Fig_intro} b). This is a relevant question in finance where $x_n$ may represent the price of a stock starting from its initial value $x_0$. The stock is deemed ``active'' till it crosses its initial value $x_0$ from above for the first time, and when this happens, the stock becomes ``bad'' and typical investors get rid of this stock from their portfolios. One can set $x_0=0$ without any loss of generality. The number of stationary points then represents the number of price reversals of this ``active'' stock. 
Such first-passage functionals, i.e., the statistical properties of observables till its first-passage time, have been well studied for the Brownian case with many applications ranging from queuing theory, directed polymers, all the way to astrophysics, e.g., in  
the study of the life-time of a comet in the solar system \cite{Satya_funct}. However we are not aware of any study of such first-passage functionals for discrete-time random walks such as L\'evy flights. Our exact results in this paper on the number of the maxima/minima till the first-passage time,  
valid for random walks with arbitrary jump distributions $\phi(\eta)$ including L\'evy flights, thus provide such an example.

It is useful to summarize our main results. We compute the full distribution of the number of minima $m$ both in the fixed $N$ ensemble as well as up to the first-passage time $N_f$.
While the distributions in the two ensembles are different, we find the striking result that each of them is universal, i.e., independent of the jump distribution $\phi(\eta)$ for all symmetric and continuous jump distributions. This includes standard random walk of finite variance jumps, as well as L\'evy flights. Remarkably, this universality holds for all values of $m$ and not just for large $m$. More precisely, for the fixed $N$ ensemble (Fig. \ref{Fig_intro} a)), we show that the distribution of the number of minima $Q(m,N)$ vanishes for $m > N/2$, while it has a nonzero value for $0 \leq m \leq N/2$ given by 
\begin{align}
\begin{split}
Q(m,N)&=\frac{1}{2^N} \frac{(N+1)!}{(N-2m)!\,(2m+1)!} \;,
\end{split}
 \label{sol:Q(m,N)}
\end{align}
valid for arbitrary symmetric and continuous $\phi(\eta)$. The universality of this result can be traced back to the fact that, for the statistics of $m$, only the signs of the jumps matter and not the actual position of the walker. It is easy to see, using the symmetry of $\phi(\eta)$, that the distribution of the number of maxima $M$ up to $N$ steps has the same expression $Q(M,N)$ with $m \to M$ in Eq. (\ref{sol:Q(m,N)}). In the large $N$ limit, $Q(m,N)$ converges to a Gaussian distribution centered at $N/4$ with a variance given by $N/16$, with non-Gaussian large deviation tails that we compute explicitly. Furthermore, we show that the joint distribution $\mathcal{Q}(m,M,N)$ of the number of minima $m$ and maxima $M$ up to step $N$ is also universal, i.e., independent of $\phi(\eta)$, for $m$, $M$ and $N$. This result also demonstrates nontrivial universal correlations between $m$ and $M$. In particular the connected two-point correlation, for $N \geq 2$, is given by
\bea \label{correl_mM}
C_c(N) = \langle m M\rangle - \langle m \rangle \langle M \rangle = \frac{N-3}{16} \;.
\eea
Interestingly they are anti-correlated for $N=2$, uncorrelated for $N=3$ and positively correlated for $N > 3$. 

For the first-passage ensemble, we show that   
the distribution $Q^{({\rm fp})}(m)$ of the number of minima $m$ till the first-passage time to the origin is also universal for all $m$ and is given by
\begin{align}
Q^{\rm (fp)}(m) = 
\begin{cases}
\frac{3}{4} ~& ~\text{for}~m=0 \;, \\
\frac{1}{2^{2m+2}}~\frac{(2m)!}{m!(m+1)!}~&~\text{for}~m\ge 1 \;.
\end{cases}
\label{sol:Q^fp(m)}
\end{align}
It turns out that the mechanism responsible for the universality in the first-passage ensemble is completely different from that of the fixed $N$ ensemble,  because here the statistics of $m$ actually depends on the position of the walk, since the position has to remain positive till the first crossing of the origin. We show that the universality in this case can be traced back, via a nontrivial mapping, to the Sparre Andersen theorem~\cite{SA} for the survival probability
of one-dimensional random walks starting at the origin. Unlike in the fixed $N$ ensemble, the distribution $Q^{\rm (fp)}(m)$ has a power-law tail $\sim m^{-3/2}$ for large $m$, indicating that all moments of $m$, including its average $\langle m \rangle$, diverge.  

We start with the fixed $N$ ensemble of the random walk defined in Eq. (\ref{def_RW}). Let us first define the ``spin'' variables $s_i = {\rm sgn}(\eta_i) = \pm 1$, which are also independent. A stationary point (a maximum or a minimum) of the random walk landscape occurs at step $i$ if $s_i s_{i+1} = -1$, irrespective of the starting point of the walk. Thus the statistics of stationary points does not depend on the actual magnitude but rather only on the signs of the jump variables $\eta_i$'s. Hence for all symmetric jump distribution $\phi(\eta)$, one expects these statistics to be universal. For instance, the total number of stationary points $K$ is just the number of ``bonds'' such that $s_i s_{i+1} = -1$. Given that there are $N-1$ bonds and each of them are equally likely to be $\pm 1$, it follows that the distribution $P(K,N)$ is simply given by the binomial distribution $P(K,N)={{N-1} \choose K}/2^{N-1}$ for $K=0,1, \cdots, N-1$ and $N \geq 2$. However, deriving the distribution of the number of minima $m$ by such a simple combinatorial argument is less trivial. To compute this distribution, we first define $Q_{\pm}(m,N)$ denoting respectively the probability of having $m$ minima in $N$ steps with the first jump either in `$+$' or `$-$' direction (irrespective of the starting position). They follow the recursion relations
\begin{align}
Q_+(m,N) &= \frac{Q_+(m,N-1)+Q_-(m,N-1)}{2} \label{Q_pm(m,N)a} \\ 
Q_-(m,N) &= \frac{Q_+(m-1,N-1)+Q_-(m,N-1)}{2} \;,
\label{Q_pm(m,N)b}
\end{align}
valid for $N \geq 3$. These recursion relations can be understood by observing what happens in the trajectory after the first jump (analogue of backward Fokker-Planck equations). If the first step is positive (which happens with probability $1/2$) and the second step is either positive or negative, no minimum is created. This explains Eq. (\ref{Q_pm(m,N)a}). In contrast, if the first step is negative, then a minimum is created if the second step is positive and hence the number of minima in the rest of the trajectory must be $m-1$. This explains the first term of Eq. (\ref{Q_pm(m,N)b}). However, if the second step is negative, no new minimum is created, explaining the second term in Eq. (\ref{Q_pm(m,N)b}). For $N=2$, it is easy to see by direct inspection that $Q_+(m,2) = \delta_{m,0}/2$ and $Q_-(m,2) = (\delta_{m,0}+ \delta_{m,1})/4$. The distribution $Q(m,N)$ of $m$ is then given by $Q(m,N) = Q_+(m,N) + Q_{-}(m,N)$.

These recursion relations (\ref{Q_pm(m,N)a})-(\ref{Q_pm(m,N)b}) can be solved using generating function techniques (see \cite{SM} for details), which leads to the result in Eq. (\ref{sol:Q(m,N)}). Note that in deriving this result we only used the symmetry of $\phi(\eta)$ but it does not have to be continuous. Indeed, for the binary jump distribution $\phi(\eta) = (\delta_{\eta,1} + \delta(\eta,-1))/2$, this result also holds. In Fig. \ref{Fig_Q(m,N)}, we verify this analytical result via numerical simulations for four \textcolor{black}{additional} different jump distributions. By expanding the factorials in Eq. (\ref{sol:Q(m,N)}) using Stirling formula, one can analyse the asymptotic scaling limit where both $m$ and $N$ are large but with their ratio $\alpha = m/N$ fixed. We find that $Q(m,N)$ takes a large deviation form
$Q(m,N) \approx e^{-N \,\Phi(\alpha=m/N)}$, 
where the rate function $\Phi(\alpha) = \ln 2 + 2 \alpha \ln(2 \alpha) + (1-2 \alpha)\ln(1-2\alpha)$, has a unique minimum at $\alpha = 1/4$. Expanding $\Phi(\alpha)$ around $\alpha=1/4$, one gets to leading order $\Phi(\alpha)\approx 8(\alpha-1/4)^2$. Substituting this quadratic behavior in the large deviation form, we get a Gaussian distribution for the typical fluctuations of $m$, with mean $\langle m \rangle \approx N/4$ and variance $N/16$. Interestingly, this limiting Gaussian distribution was derived for the special case of Bernoulli random walk with $\phi(\eta) = (\delta_{\eta,1} + \delta_{\eta,-1})/2$ in the maths literature by a different method~\cite{Marckert}, but the issue of the universality of $Q(m,N)$ for all $m$ and $N$ was not noticed. 

Since $\phi(\eta)$ is symmetric, we can reflect the trajectory $x_n \to -x_n$ such that the local minima become the local maxima. Hence, one finds that the distribution of the number of maxima $M$ is again given by the same result in Eq. (\ref{sol:Q(m,N)}) with $m$ replaced by $M$. What is however more interesting is to investigate if there are correlations between $m$ and $M$. To characterize these correlations, we define ${\cal Q}_{\pm}(m,M,,N)$ as the joint distribution of $m$ and $M$ in $N$ steps, starting with a positive or negative jump respectively. As in the case of $Q(m,N)$, this joint distribution is also independent of the starting point $x_0$ for symmetric $\phi(\eta)$. Following the same steps as in Eqs. (\ref{Q_pm(m,N)a})-(\ref{Q_pm(m,N)b}) by counting what happens after the first jump, we can write down a pair of exact recursion relations (see~\cite{SM} for details). The 
triple generating function of the joint distribution ${\cal Q}(m,M,,N) = {\cal Q}_{+}(m,M,,N)+ {\cal Q}_{-}(m,M,,N)$ with respect to $m, M$ and $N$, defined by $S(u,v,z) = \sum_{m \geq 0, M\geq 0, N \geq 2} {\cal Q}(m,M,,N) u^m v^M z^N$, 
can be computed explicitly as \cite{SM}
\begin{eqnarray}
S(u,v,z) =  z^2 \frac{(2-z)+(u+v)+zuv}{(2-z)^2-uvz^2} \label{sol:S(v,u,z)} \;.
\end{eqnarray}
Clearly this result is again universal, i.e., independent of $\phi(\eta)$. From this formula, one can derive all moments by taking derivatives with respect to $u$ and $v$. For example, for the two-point connected correlation function $C_c(N) = \langle m M \rangle - \langle m \rangle \langle M \rangle$, we get the exact result given in Eq.~(\ref{correl_mM}). In \cite{SM} we compare our analytical prediction with numerical simulations for \textcolor{black}{five} different jump distributions, finding excellent agreement. 

\begin{figure}
\centering
\includegraphics[width = 0.98\linewidth]{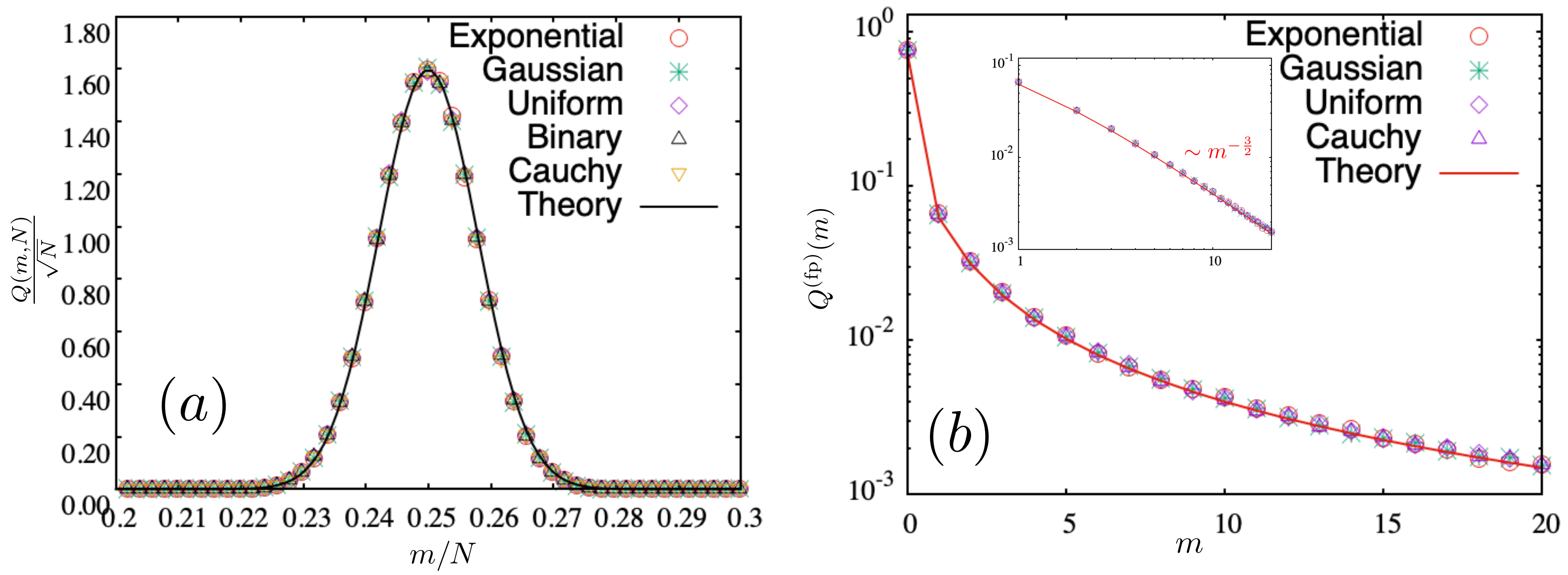}
\caption{{\bf a)} The universal expression of  $Q(m,N)$ in Eq.~\eqref{sol:Q(m,N)} is verified numerically for \textcolor{black}{five} choices of $\phi(\eta)$: (i) Exponential: $\phi(\eta) = \frac{a }{2}\exp(-a |\eta|)$ (ii) Gaussian: $\phi(\eta) = \frac{1 }{\sqrt{2\pi\sigma^2}}\exp\left(-\frac{\eta^2}{2 \sigma^2}\right)$, (iii) Uniform: $\phi(\eta) =\frac{1}{2\Delta}\Theta\left(z+\Delta\right)\Theta\left(\Delta-z\right)$, (iv) Binary: $\phi(\eta) =\frac{1}{2}\left[ \delta(\eta-1) + \delta(\eta+1)\right]$ \textcolor{black}{and (v) Cauchy: $\phi(\eta) = a/(\pi(\eta^2+a^2))$}. {\bf b)} Numerical results for the distribution $Q^{\rm (fp)}(m)$ plotted as function for $m$ for \textcolor{black}{four} different symmetric and continuous jump distributions. Also plotted the analytical result of $Q^{\rm (fp)}(m)$ in Eq.~\eqref{sol:Q^fp(m)} by a solid line. The fact that they coincide for all $m$
indicates the strong universality of the result. The inset shows numerical verification of the $m^{-3/2}$ decay of $Q^{\rm (fp)}(m)$ for large $m$. \textcolor{black}.} 
\label{Fig_Q(m,N)}
\end{figure}

We now turn to the calculation of the distribution of the number of minima $Q^{\rm{(fp)}}(m)$ in the first-passage ensemble. Unlike in the fixed $N$ ensemble, here the translation invariance is lost since the process stops when it hits the origin for the first time. Here it is convenient to define 
 $Q_{\pm}^{\rm{(fp)}}(x_0,m)$ as the probability to have $m$ minima till the first-passage, starting from the initial position $x_0$ and with initial step positive or negative. One can write down the backward recursion relations satisfied by $Q_{\pm}^{\rm{(fp)}}(x_0,m)$, again by observing what happens in the first step. However, unlike in the fixed $N$ ensemble, where these equations were independent of the noise distribution (see Eqs. \eqref{Q_pm(m,N)a}-\eqref{Q_pm(m,N)b}), for the first-passage ensemble, the recursion relations are integral equations that explicitly involve $\phi(\eta)$. These equations are a bit too long to display here, hence we present them in detail in the Supp. Mat. \cite{SM}. It turns out that, for generic $\phi(\eta)$, it is hard to solve these integral equations. However, for the special case $\phi(\eta) = (1/2)\, e^{-|\eta|}$, they are exactly solvable as shown in~\cite{SM}. In this special case, setting $x_0=0$, we find the result for $Q^{\rm{(fp)}}(m) = Q_{+}^{\rm{(fp)}}(x_0=0,m)+ Q_{-}^{\rm{(fp)}}(x_0=0,m)$ given in Eq. (\ref{sol:Q^fp(m)}). Then we performed numerical simulations for other continuous and symmetric jump distributions $\phi(\eta)$ (not necessarily double-exponential) and, amazingly, the simulation points fell exactly on top of the results (\ref{sol:Q^fp(m)}) for the double-exponential jump distribution (see Fig. \ref{Fig_Q(m,N)} b)). This indicated that the result in Eq. (\ref{sol:Q^fp(m)}) is also universal for all $m$. Such a strong universality (for all $m$) came as an unexpected surprise and the mechanism behind it is far from obvious. To understand this universality, below we first map the minima counting problem to an auxiliary discrete-time random walk problem with an effective jump distribution. Under this mapping, the distribution $Q^{\rm{(fp)}}(m)$ in the original problem is related exactly to the survival probability of this auxiliary walk up to step $m$. Then, using the universality of the latter quantity via the celebrated Sparre Andersen theorem \cite{SA}, we prove this amazing~universality. Thus the mechanism behind this universality in the first-passage ensemble is much more subtle than the universality encountered before in the fixed time ensemble.

\begin{figure}[t]
\centering
    \includegraphics[width = \linewidth]{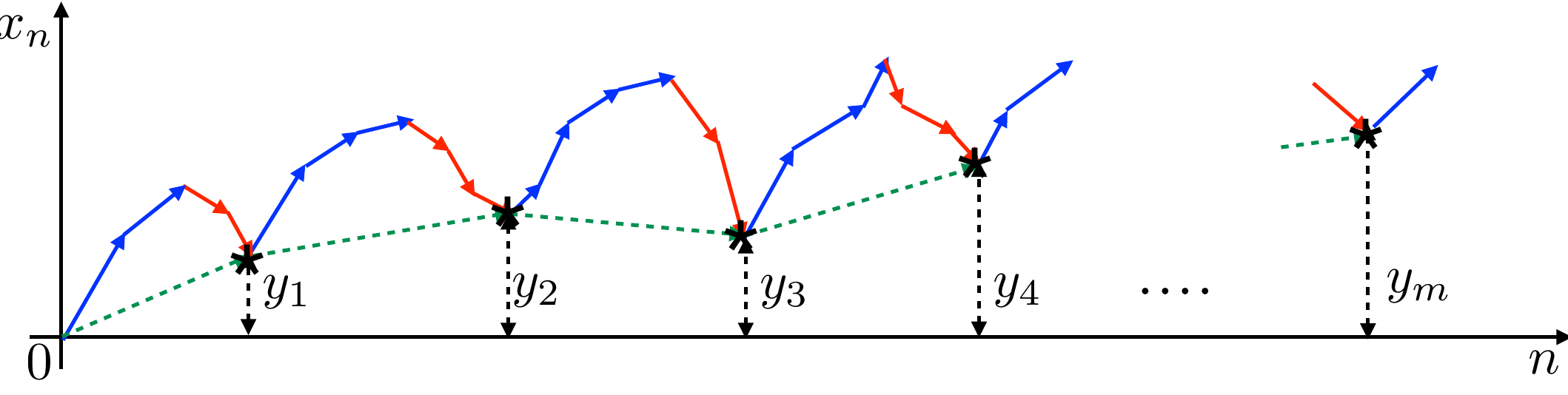}
\caption{Schematic trajectory of the original random walk where the successive minima are shown by stars with heights $y_k$, with $k=1,2, \cdots$. Note that, by construction, there is a single peak between two successive minima, consisting of monotonically increasing jumps on the left of the peak (shown by blue arrows) and monotonically decreasing jumps on the right (red arrows). The $y_k$'s constitute an auxiliary random walk, $y_k = y_{k-1}+\xi_k$ with an effective jump distribution $\Psi(\xi)$ that is continuous and symmetric.}\label{Fig_traj}    
\end{figure}

To construct this mapping, we consider a typical trajectory of the original random walk which has at least $m$ minima as shown in Fig. \ref{Fig_traj}. Let $\{y_i\}$'s denote the heights of these minima. Then the first crucial point to realise is that between any two successive minima, there can be only one global maximum with monotonically increasing jumps on its left (shown by blue arrows) and monotonically decreasing jumps on its right (red arrows). This is because if there is more than one peak between the two minima, that would automatically mean that there is an additional minimum between the two, which is ruled out by construction since we are considering successive minima. The height $y_k$ of the $k$-th local minimum can be expressed as $y_k = y_{k-1} + \xi_k$ where $\xi_k$ represents the difference in heights between the $(k-1)$-th and the $k$-th minima. Thus $y_k$ represents the position of the auxiliary random walk at step $k$, starting from $y_0=0$.

The next step is to compute the distribution $\Psi(\xi)$ of the jump variable $\xi_k$, which can be computed explicitly in terms of the original jump distribution $\phi(\eta)$ (see \cite{SM}). However, the detailed form of $\Psi(\xi)$ is not needed to prove the universality. What is needed is to show that this jump distribution $\Psi(\xi)$ is continuous and symmetric, which can be proved rigorously as detailed in \cite{SM}. The last important step is to realise that the cumulative probability of having at least $m$ minima in the trajectory is exactly the event $\{y_1 \geq 0, y_2 \geq 0, \cdots, y_m \geq 0\}$ in the auxiliary random walk. In addition, one needs to ensure that the last position $y_m$ is actually a local minimum, which happens with probability $1/2$. Thus we have the exact relation
\bea \label{relation}
\sum_{k=m}^\infty Q^{\rm{(fp)}}(k) = \frac{1}{2} q_m \;,
\eea  
where $q_m$ is the probability that the auxiliary walk $y_k$ stays non-negative up to step $m$ and the factor $1/2$
comes from the fact that the $m$-th position of the auxiliary walk must be a local minimum of the original walk.
The survival probability $q_m$ of the auxiliary walk up to step $m$, starting at the origin, is simply universal, i.e.,
independent of the effective jump distribution $\Psi(\xi)$ as long as it is symmetric and continuous. It 
is given by the Sparre Andersen formula $q_m = {2m \choose m}\, 2^{-2\, m}$, for $m=0,1,2,\ldots$~\cite{SA}. Hence, using Eq.~(\ref{relation}), it follows that the probability of having exactly $m$ minima up to the first-passage time, is given, for $m\ge 1$, by 
the universal formula
\begin{equation}
Q^{\rm{(fp)}}(m)= \frac{1}{2}\, [q_m-q_{m+1}]= \frac{1}{2^{2m+2}}\, \frac{(2m)!}{m!\, (m+1)!}\, \label{prob_nmin.1} \;.
\end{equation}
The special case $m=0$ has to be dealt with separately (see \cite{SM}), leading again to the universal result $Q^{\rm{(fp)}}(0)=3/4$ as in Eq.~(\ref{sol:Q^fp(m)}).   

To conclude, we have shown that the distribution of the number of minima/maxima of
a random walk landscape is universal, i.e., independent of the jump distribution. We have computed this 
distribution exactly both for a fixed number of steps as well as till the first-passage time. These universal results
are valid even for long-ranged landscapes generated by L\'evy flights. Indeed, for the L\'evy flights, our result provides
a rare exactly solvable example of a first-passage functional. 
Our results can be directly applied to the landscape generated by an RTP (see \cite{SM}). For instance, one can show that the number of stationary points of the RTP landscape is simply a Poissonian with mean $\gamma t/2$, independent of the post-tumble velocity distribution. The distribution of the number of minima for an RTP of duration $t$ is also universal, but highly non-Poissonian as shown in \cite{SM}.  
Our work opens up many interesting directions. For example, it would be interesting to compute the distribution of minima/maxima for landscapes generated by anomalous sub-diffusive processes~\cite{BG,Metzler,Godec}.

\sectionprl{Acknowledgements}
AK and SNM would like to thank the Isaac Newton Institute for Mathematical Sciences, Cambridge, for support and hospitality during the programme {\it New statistical physics in living matter: non equilibrium states under adaptive control} where work on this paper was undertaken. This work was supported by EPSRC Grant Number EP/R014604/1. SNM and GS acknowledge support from ANR Grant No. ANR-23-CE30-0020-01 EDIPS. \textcolor{black}{AK would like to acknowledge the support of DST, Government of India Grant under Project No. ECR/2017/000634 and the MATRICS 
grant MTR/2021/000350 from the SERB, DST, Government of India. AK acknowledges the Department of Atomic Energy, 
Government of India, for their support under Project No. RTI4001. }.


%

\end{document}


\title{\titlename}

%

\author{Anupam Kundu}
\address{International Centre for Theoretical Sciences, Tata Institute of Fundamental Research, Bengaluru -- 560089, India}
\author{Satya N. Majumdar}
\address{LPTMS, CNRS, Univ. Paris-Sud, Universit\'e Paris-Saclay, 91405 Orsay, France}
\author{Gr\'egory Schehr}
\address{Sorbonne Universit\'e, Laboratoire de Physique Th\'eorique et Hautes Energies, CNRS UMR 7589, 4 Place Jussieu, 75252 Paris Cedex 05, France}	
	
\date{\today}
	
\begin{abstract}
We provide some details of the derivation of the results presented in the main text. 
\end{abstract}	
	
\maketitle

\section{Distribution of the number of minima/maxima of a random walk landscape of fixed number of steps $N$}

In this section we provide the derivation of the main results for the fixed $N$ ensemble of the random walk landscape. In subsection \ref{sub:minima}, we provide the derivation of the universal result for $Q(m,N)$ denoting the distribution of the number of minima up to step $N$. In subsection \ref{sub:min_max} we derive the joint distribution ${\cal Q}(m,M,N)$ of the number of minima $m$ and the number of maxima $M$ up to $N$ steps. In subsection \ref{total}, we provide an independent derivation of the distribution of the total number of stationary points $P(K,N)$ up to step $N$ where $K=m+M$. In subsection \label{sec:RTP}, we provide a direct application of our results for random walk to a landscape generated by a run-and-tumble particle~(RTP).

\subsection{Distribution $Q(m,N)$ of the number of minima $m$ up to step $N$}\label{sub:minima}

\begin{figure}[h]
\includegraphics[width = 0.7\linewidth]{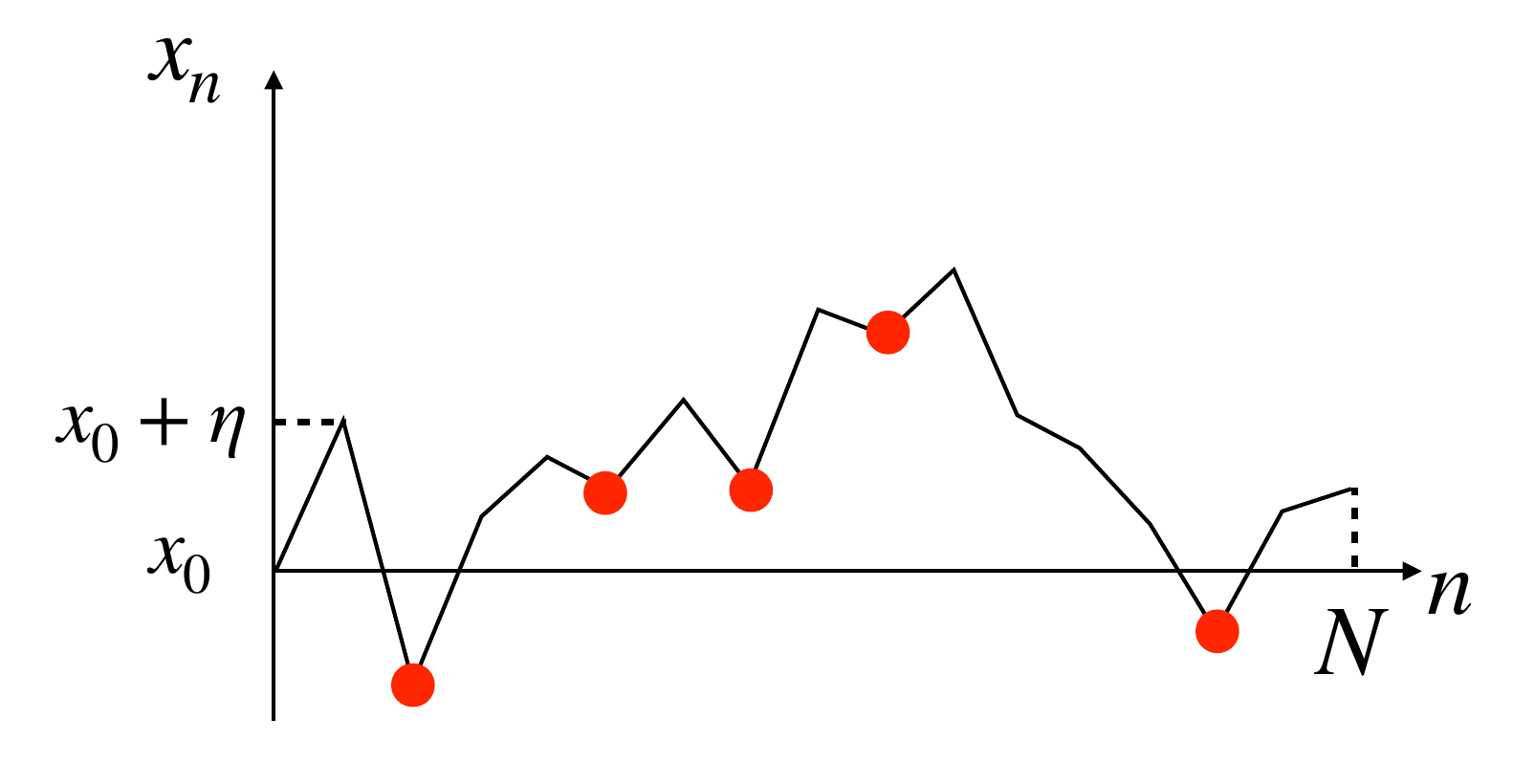}
\caption{A trajectory of a random walk of $N$ steps starting at $x_0$. At the first step the walker jumps to $x_0+\eta$ where $\eta$ is a random jump drawn from $\phi(\eta)$ which is symmetric and continuous. The local minima are marked as filled red circles.}\label{Fig_SM_1}
\end{figure}

Here we provide the derivation of the recursion relations in Eqs. (5)-(6) together with its solution in Eq. (2) of the main text.
We start with a discrete-tile random walk evolving on a continuous line with the position $x_n$ at step $n$ updated via the Markov jump rule (see Fig. \ref{Fig_SM_1})
\bea \label{def_RW_SM}
x_n = x_{n-1} + \eta_n \;,
\eea
where $\eta_n$'s are independent and identically distributed (IID) random variables, each drawn from a continuous and symmetric distribution $\phi(\eta)$. The walker starts at $x_0$. To compute the distribution of the number of minima up to step $N$, it is convenient to introduce a pair of quantities $Q_{\pm}(x_0,m,N)$ denoting respectively the probability of having $m$ minima in $N$ steps, starting from $x_0$ with the first jump either in `$+$' or `$-$' direction. The idea is to write down an exact pair of recursion relations by observing what happens after the first jump. This is the analogue of backward Fokker-Planck equations. The pair of recursion relations read 
\begin{widetext}
\bea
&&Q_+(x_0,m,N) = \int_0^\infty d\eta\, \left[ Q_+(x_0+\eta,m,N-1) + Q_-(x_0+\eta,m,N-1)\right] \phi(\eta)  \label{FP_1} \\
&&Q_-(x_0,m,N) = \int_{-\infty}^0  d\eta\, \left[ Q_+(x_0+\eta,m-1,N-1) + Q_-(x_0+\eta,m,N-1)\right] \phi(\eta) \label{FP_2} \;,
\eea 
\end{widetext}
where $\eta$ denotes the first random jump. If the walker starts with a positive (or negative) jump, it arrives at the next step at $x_0+\eta$ with $\eta \geq 0$ (respectively $\eta \leq 0$) -- see Fig. \ref{Fig_SM_1}. In the case when the first jump is positive, there is no new minimum generated by the first jump and hence in the recursion $m$ remains the same in Eq. (\ref{FP_1}). In contrast, if the first jump is negative and the second one is positive, it creates a minimum at the end of the first step. This means that we need to have $m-1$ minima for the rest of the $N-1$ steps, starting at $x_0+\eta$ with $\eta \leq 0$. This explains the first term in Eq. (\ref{FP_2}). Similarly, if both the first and the second jumps are negative, there is no minimum at the end of the first step and this leads to the second term in Eq.~(\ref{FP_2}). Fortunately, one can exploit the translational invariance with respect to the initial position $x_0$, i.e., the fact that $Q(x_0,m,N)$ is actually independent of $x_0$. This is because, no matter where the random walk starts, it is only the relative signs of the jumps that can create a minimum (a negative jump followed by a positive jump). Using the independence of $Q_{\pm}(x_0,m,N) \equiv Q_{\pm}(m,N)$ on $x_0$ and the symmetry of the jump distribution namely $\int_0^\infty \phi(\eta) d\eta = \int_{-\infty}^0 \phi(\eta)\, d \eta = 1/2$, the recursion relations (\ref{FP_1})-(\ref{FP_2}) simplify to
\begin{align}
Q_+(m,N) &= \frac{Q_+(m,N-1)+Q_-(m,N-1)}{2} \label{Q_pm(m,N)a} \\ 
Q_-(m,N) &= \frac{Q_+(m-1,N-1)+Q_-(m,N-1)}{2} \;,
\label{Q_pm(m,N)b}
\end{align}
valid for $N \geq 3$. For $N=2$, it is easy to show that $Q_+(m,2) = \delta_{m,0}/2$ and $Q_-(m,2) = (\delta_{m,0}+ \delta_{m,1})/4$. The distribution $Q(m,N)$ of $m$ is then given by $Q(m,N) = Q_+(m,N) + Q_{-}(m,N)$. These recursion relations (\ref{Q_pm(m,N)a})-(\ref{Q_pm(m,N)b}) already demonstrate that the dependence on the noise distribution has dropped out, indicating that $Q_{\pm}(m,N)$, and consequently $Q(m,N)$, 
are universal for all $m$ and $N$. 

In order to solve the coupled recursion relations \eqref{Q_pm(m,N)a} and \eqref{Q_pm(m,N)b}, we define the following generating functions 
\begin{align}
Z_\pm(m,z) = \sum_{N=2}^\infty Q_\pm(m,N)z^N,~~~|z| < 1 \;. \label{Z_pm(m,z)}
\end{align}
Multiplying both sides of Eq.~\eqref{Q_pm(m,N)a} and \eqref{Q_pm(m,N)b} by $z^N$ and summing over $N\ge 3$, one finds
\begin{eqnarray}
Z_+(m,z) &=& \frac{z}{2} \left[ Z_+(m,z) + Z_-(m,z)\right] + \frac{z^2}{2}\delta_{m,0},~~\text{for}~~m\ge 0 \;, \label{Z_+(m,z)} \\
Z_-(m,z) &=& \frac{z}{2} \left[ Z_+(m-1,z) + Z_-(m,z)\right] + \frac{z^2}{4}[\delta_{m,0}+\delta_{m,1}],~~\text{for}~~m\ge 0 \;, \label{Z_-(m,z)} 
\end{eqnarray}
with the convention $Z_+(-1,0)=0$. Setting $m=0$ in the second equation gives
\bea \label{Zm0}
Z_-(0,z) = \frac{z^2}{2(2-z)} \;.
\eea
Substituting this in the first line with $m=0$ then gives
\bea \label{Zp0}
Z_+(0,z) = \frac{z^2(4-z)}{2(2-z)^2} \;.
\eea

\noindent
To solve Eqs.~\eqref{Z_+(m,z)} and \eqref{Z_-(m,z)} we further define the following generating functions with respect to $m$ 
\begin{align}
S_\pm(u,z) = \sum_{m=0}^\infty Z_\pm(m,z)u^m = \sum_{N=2}^\infty \sum_{m=0}^\infty u^mz^NQ_\pm(m,N) \;. \label{S_pm(u,z)}
\end{align}
Multiplying both sides of Eqs.~\eqref{Z_+(m,z)} and \eqref{Z_-(m,z)} by $u^m$ and summing over $m\ge 0$, we get 
\begin{eqnarray}
S_+(u,z) &=&  \frac{z}{2} \left[ S_+(u,z) + S_-(u,z)\right] + \frac{z^2}{2}. \label{S_+(u,z)} \\
S_-(u,z) &=&   \frac{z}{2} \left[ uS_+(u,z) + S_-(u,z) \right] + \frac{z^2}{4}(1+u) \;.
\end{eqnarray}
Solving this pair of equations gives
\begin{align}
S_+(u,z) &= \frac{z^2[4+(u-1)z]}{2[(2-z)^2-uz^2]} \;, \label{sol:S_+(u,z)} \\
S_-(u,z) &= \frac{z^2[2(1+u)-z(1-u)]}{2[(2-z)^2-uz^2]} \;. \label{sol:S_-(u,z)} 
\end{align}
The sum $S(u,z) = S_+(u,z)+S_-(u,z)$ is given by
\bea \label{SumS}
S(u,z) = \frac{z^2(3+u-z+u z)}{(2-z)^2 - uz^2} \;.
\eea

Inverting Eq. (\ref{sol:S_+(u,z)}) using Cauchy's formula, one finds
\bea
Z_+(m,z) = \frac{1}{2 \pi i} \oint_{C_0} du \frac{1}{u^{m+1}} \frac{1}{2} \frac{4+(u-1)z}{\left[ \left( \frac{2-z}{z}\right)^2 -u\right]} \;, \label{cauchy-n} 
\eea
where $C_0$ is a contour around $u=0$ in the complex $u$-plane. Noting that the integrand has a simple pole at $u = ((2-z)/z)^2$, this integral can be trivially done to give 
\begin{figure}
\centering
\includegraphics[scale=0.2]{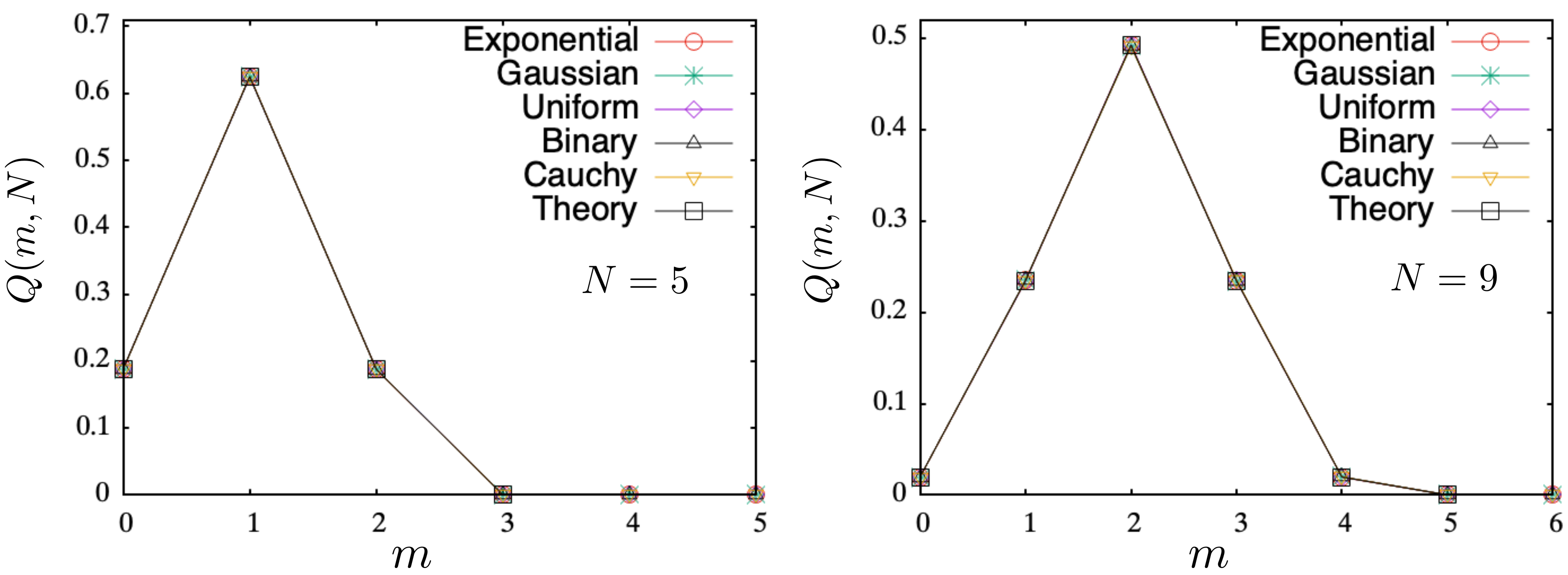}
\caption{Plot of the numerically obtained $Q(m,N)$ for five different jump distributions and for two different values of $N$, compared to theoretical expression in Eq.~\eqref{sol:Q(m,N)}, plotted with square symbols, showing a perfect agreement. The collapse of the data for the different jump distributions on a single curve
clearly demonstrates  the universality of $Q(m,N)$.}
\label{QmN-5}
\end{figure}
\bea
Z_+(m,z)
 = \begin{cases}
\frac{2z^{2m+1}}{(2-z)^{2m+2}},~&~\text{for}~m \ge 1 \;, \\
 & \\ 
\frac{z^2(4-z)}{2(2-z)^{2}},~&~\text{for}~m=0 \;.
\end{cases}
\label{sol:Z_+(m,z)}
\eea
Similarly, for $Z_-(m,z)$ we get
\begin{align}
Z_-(m,z) = 
\begin{cases}
\frac{z^{2m}}{2(2-z)^{2m+1}},~&~\text{for}~m \ge 1 \;, \\
 & \\
\frac{z^2}{2(2-z)},~&~\text{for}~m=0 \;.
\end{cases}
\label{sol:Z_-(m,z)}
\end{align}
Expanding further in powers of $z$ gives the desired results
\begin{align}
Q_+(m,N) = \frac{1}{2^N} \frac{\Gamma(N+1)}{\Gamma(N-2m)\Gamma(2m+2)},~~\text{for}~m\ge 1 \;
 \label{sol:Q_+(m,N)-mge-1}
\end{align} 
and 
\begin{align}
Q_-(m,N)&=\frac{1}{2^N} \frac{\Gamma(N+1)}{\Gamma(N+1-2m)\Gamma(2m+2)},~~~~\text{for}~~0\leq m \leq \frac{N}{2} \;. 
\label{sol:Q_-(m,N)}
\end{align}
Finally, $Q(m,N)=Q_+(m,N)+Q_-(m,N)$ is given by 
\begin{align}
Q(m,N)&=\frac{1}{2^N} \frac{\Gamma(N+2)}{\Gamma(N+1-2m)\Gamma(2m+2)},~~\text{for}~~0\leq m \leq \frac{N}{2} \;.
 \label{sol:Q(m,N)}
\end{align}
This result is universal for all $m$ and $N$ and is valid for any arbitrary jump distribution $\phi(\eta)$ as long as $\phi(\eta)$ is symmetric and continuous. 
We have verified this universal result for $Q(m,N)$ numerically for two different values of $N$ and for different jump distributions as shown in Fig. \ref{QmN-5} and also in Fig. 1 of the main text. From this formula (\ref{sol:Q(m,N)}) one can calculate all the moments of $m$. For example, the mean and the variance are given by
\bea \label{mean_varm}
\langle m \rangle = \frac{N-1}{4} \quad, \quad {\rm Var}(m) = \langle m^2 \rangle - \langle m \rangle^2 = \frac{N+1}{16} \;.
\eea

\subsection{Joint distribution ${\cal Q}(m,M,N)$ of the number of minima $m$ and the number of maxima $M$ up to step $N$}\label{sub:min_max}

In this subsection we study the joint probability distribution, denoted by ${\cal Q}(m,M,N)$, of having $m$ minima and $M$ maxima up to step $N$ of the random walk in Eq. (\ref{def_RW_SM}). As in the case of the distribution of the number of minima discussed in the previous subsection, the joint distribution ${\cal Q}(m,M,N)$ is also independent  of the starting point $x_0$ of the walk. As in the previous section, it is convenient to define ${\cal Q}_\pm(m,M,N)$ denoting the joint distribution of $m$ and $M$ with the first step positive or negative. Then ${\cal Q}(m,M,N)={\cal Q}_+(m,M,N)+{\cal Q}_-(m,M,N)$. By investigating what happens after the first jump, it is straightforward to write down the pair of backward recursion relations 
\begin{eqnarray}
{\cal Q}_+(m,M,N) &= \frac{1}{2} \left[ {\cal Q}_+(m,M,N-1)+{\cal Q}_-(m,M-1,N-1)\right], \\ 
{\cal Q}_-(m,M,N) &= \frac{1}{2} \left[ {\cal Q}_+(m-1,M,N-1)+{\cal Q}_-(m,M,N-1)\right], 
\label{Q_pm(M,N)} 
\end{eqnarray}
which are valid for $N \geq 3, M \geq 0$ and $m \geq 0$ with the convention
\begin{eqnarray}
{\cal Q}_+(m,-1,N)&=0,~~{\cal Q}_-(-1,M,N)=0,~~\text{for}~N\ge 0 \;. 
\end{eqnarray}
For $N=2$, it is easy to show that
\begin{align}
{\cal Q}_+(m,M,2)&=\frac{1}{4}\left[ \delta_{M,1} + \delta_{m,0}\right], \label{Q_+(M,m,2)} \\
{\cal Q}_-(m,M,2)&=\frac{1}{4}\left[ \delta_{M,0} + \delta_{m,1}\right]. \label{Q_-(M,m,2)} 
\end{align}
The generating functions 
\begin{align}
Z_\pm(m,M,z) = \sum_{N=2}^\infty {\cal Q}_\pm(m,M,N)~z^N, \label{def:Z_pm}
\end{align}
then satisfy
\begin{align}
Z_+(m,M,N)&= \frac{z}{2} \left[ Z_+(m,M,z) + Z_-(m,M-1,z)\right] + \frac{z^2}{4}\delta_{m,0}[\delta_{M,1}+\delta_{M,0}], \label{eq:Z_+}\\
Z_-(m,M,N)&= \frac{z}{2} \left[ Z_+(m-1,M,z) + Z_-(m,M,z)\right] + \frac{z^2}{4}\delta_{M,0}[\delta_{m,1}+\delta_{m,0}], \label{eq:Z_-}
\end{align}
with $Z_+(-1,M,z)=0$ and $Z_-(m,-1,z)=0$. 

\begin{figure}
\centering
    \includegraphics[width=0.5\linewidth]{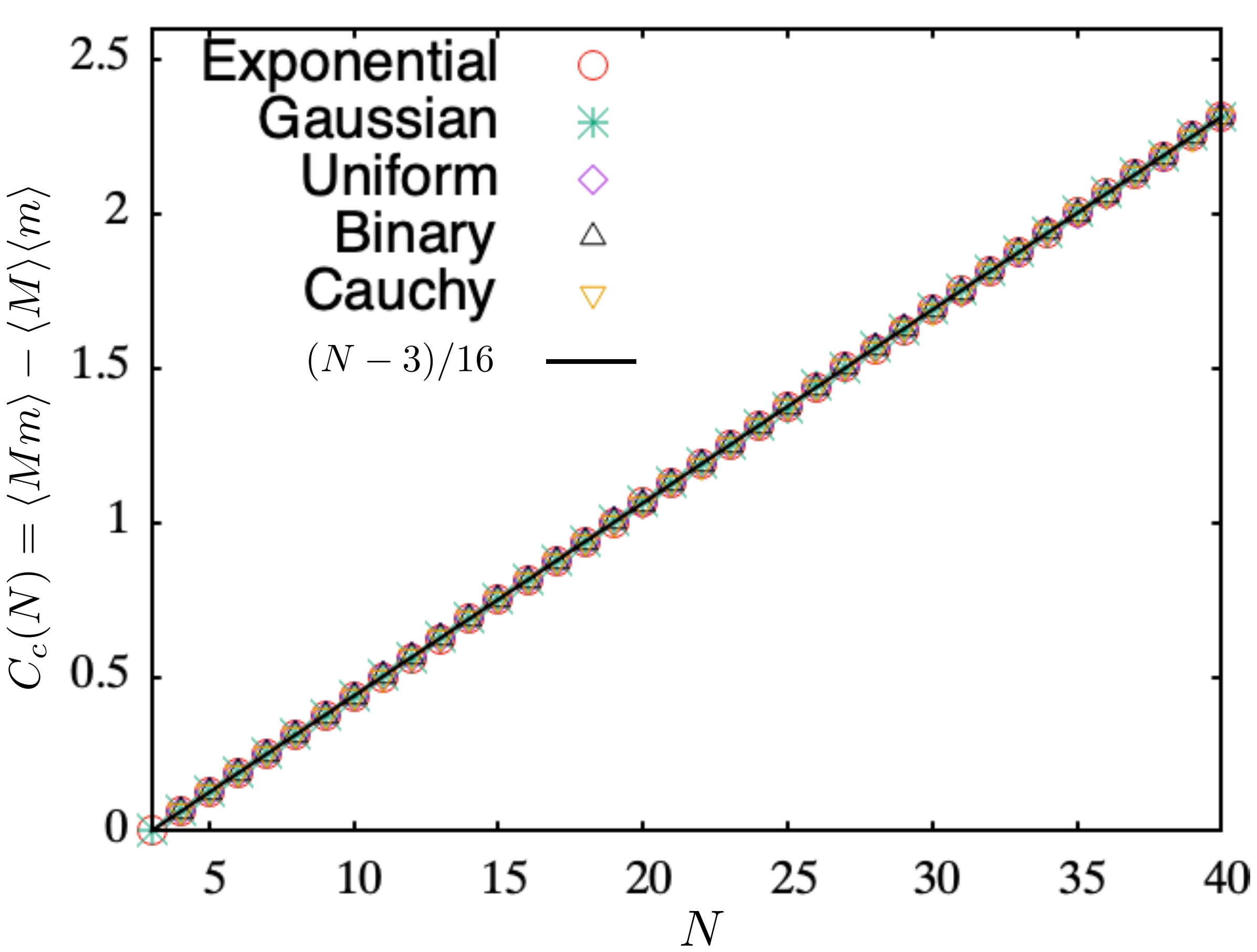}
\caption{Connected correlation function between the number of minima $m$ and the number of maxima $M$ in a random walk landscape of $N$ steps. The points represent numerical data for five different jump distributions and the solid line represents the analytical result in Eq.~\eqref{tildeC}.}
\label{fig:CMm-n}
\end{figure}

To solve Eqs. \eqref{eq:Z_+} and \eqref{eq:Z_-} we further define the following generating functions
\begin{align}
S_\pm(u,v,z) = \sum_{M=0}^\infty \sum_{m=0}^\infty Z_\pm(m,M,z)~u^m~v^M \label{def:S_pm}
\end{align}
that simply satisfy the coupled equations 
\begin{align}
\begin{split}
(2-z)S_+(u,v,z) - vz S_-(u,v,z) &= \frac{z^2}{2}(1+v) \;, \\
zuS_+(u,v,z) - (2-z) S_-(u,v,z) &= -\frac{z^2}{2}(1+u) \;.
\end{split}
\label{eq:S_pm}
\end{align}
Solving this pair of linear equations we get
\begin{align}
\begin{split}
S_+(u,v,z) & = \frac{z^2}{2} \frac{(2-z)(1+v)+zv(1+u)}{(2-z)^2-uvz^2}, \\
S_-(u,v,z) & = \frac{z^2}{2} \frac{(2-z)(1+u)+zu(1+v)}{(2-z)^2-uvz^2}.
\end{split}
\label{sol:S_pm}
\end{align}
Hence the sum $S(u,v,z) =S_+(u,v,z) +S_-(u,v,z)$ is given by
\begin{align}
S(u,v,z)=  z^2 \frac{(2-z)+(u+v)+zuv}{(2-z)^2-uvz^2} \;. \label{sol:S(v,u,z)}
\end{align}
This provides the derivation of the result in Eq. (7) of the main text. Note that by setting $v=1$ in (\ref{sol:S(v,u,z)}) one recovers the marginal generating function $S(u,z)$ in Eq. (\ref{SumS}). 

From the explicit generating function in Eq. (\ref{sol:S(v,u,z)}), one can compute different moments and correlations of $m$ and $M$. For example, let $C(N)= \langle mM\rangle$ denote the correlation between $M$ and $m$ then 
$\bar{C}(z)=\sum_{N=2}^\infty C(N)z^N$ can be obtained from $S(u,v,z)$ as 
\begin{eqnarray}
\bar{C}(z) = \left(\frac{d^2}{dudv}S(u,v,z) \right )_{{u=1},{v=1}}= \frac{z^3(2-z)}{8(1-z)^3}= \sum_{N=2}^\infty \frac{(N+1)(N-2)}{16}~z^N  
\end{eqnarray}
This gives
\begin{align}
C(N) = \frac{(N+1)(N-2)}{16},~~\text{for}~N\ge 2 \;. \label{sol:C(N)}
\end{align}
So the connected correlation $C_c(N) = \langle mM\rangle - \langle m \rangle \langle M \rangle$ is given by 
\begin{align}
C_c(N) = \frac{N-3}{16},~~\text{for}~N\ge 3, \label{tildeC}
\end{align}
where we have used $\langle m \rangle =\langle M \rangle = (N-1)/4$. This provides the derivation of the result announced in Eq. (3) in the main text. 
This result in Eq.~\eqref{tildeC} is verified numerically in Fig.~\ref{fig:CMm-n}.

\subsection{Distribution of the number of stationary points $P(K,N)$ for a random walk landscape up to $N$ steps}\label{total}

\begin{figure}[t]
\includegraphics[width=0.5\linewidth]{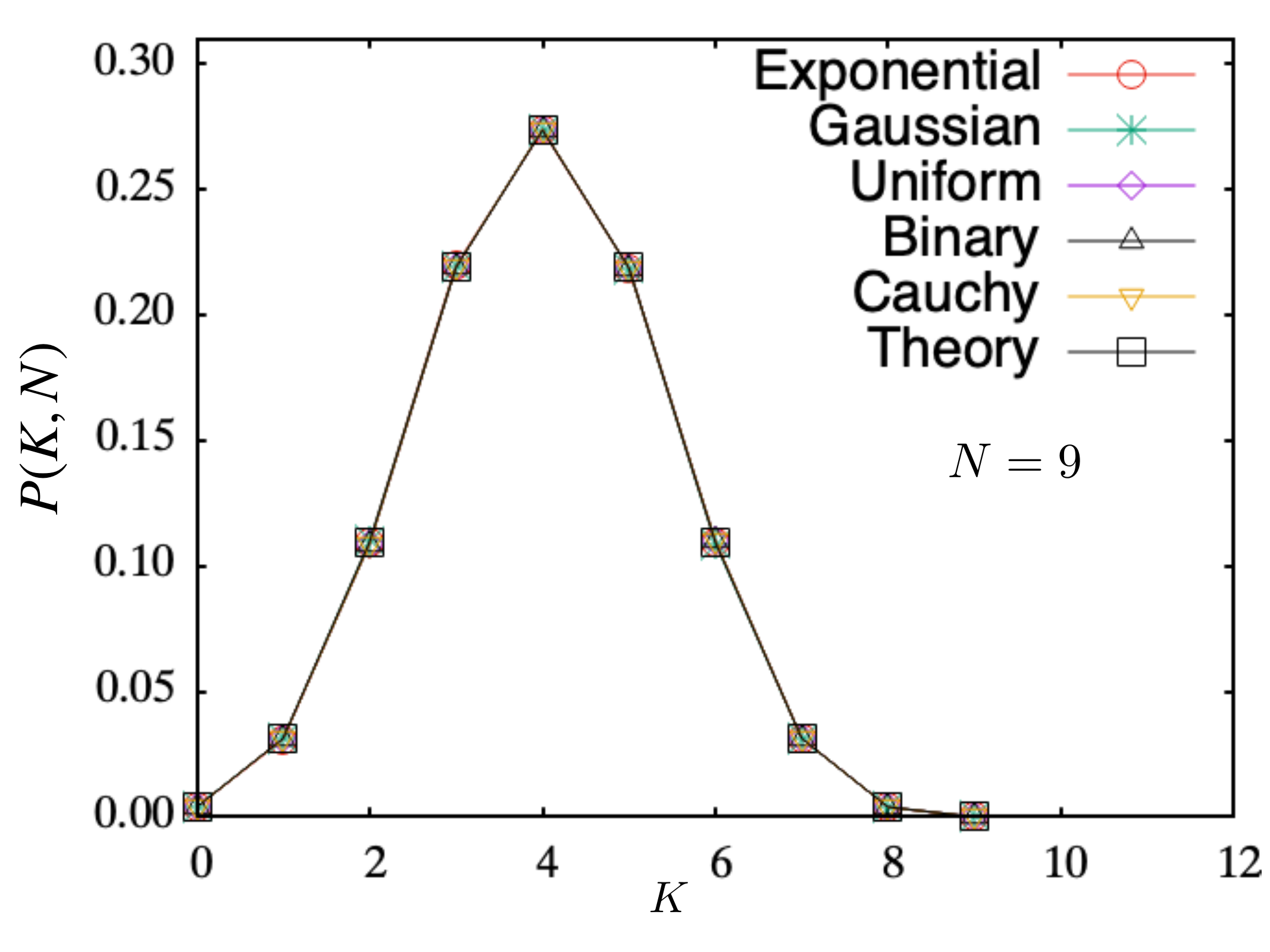}
\caption{Plot of the numerically obtained $P(K,N)$ for five different jump distributions and for $N=9$, compared to theoretical expression in Eq.~\eqref{PkN}, plotted with square symbols, showing a perfect agreement. The collapse of the data for the different jump distributions on a single curve
clearly demonstrates  the universality of $P(K,N)$.}\label{FigPkN}
\end{figure}

Let $K=m+M$ denote the total number of stationary points (maxima and minima) for a random walk landscape of $N$ steps. For the distribution $P(K,N)$ of $K$, we have provided, in the main text, a very simple combinatorial proof of the exact formula 
\bea \label{PkN}
P(K,N) = \frac{1}{2^{N-1}} {N-1 \choose K} \;, \; K = 0, 1, \cdots, N-1 \quad {\rm and} \quad N \geq 2 \;.
\eea
Here, we provide an alternative derivation of this result using the backward recursion relations as in the derivation of $Q(m,N)$, i.e., the distribution of the number of minima up to $N$ steps. Once again, the distribution $P(K,N)$ is independent of the starting position $x_0$. As usual, it is convenient to define the pair of probabilities $P_{\pm}(K,N)$ denoting the distributions starting with a positive or negative step. Investigating what happens in the first step, it is easy to see that they satisfy the recursion relations
\bea \label{FP_P}
P_+(K,N) &=& \frac{1}{2} P_{+}(K,N-1) + \frac{1}{2} P_-(K-1,N-1) \label{Pp}\\
P_-(K,N) &=&\frac{1}{2} P_{-}(K,N-1) + \frac{1}{2} P_+(K-1,N-1) \;, \label{Pm}
\eea
valid for $N \geq 3$ and $K\geq 1$. Since $P(K,N) = P_+(K,N) + P_-(K,N)$, it follows by adding these two equations that the recursion relation for the sum is closed and reads
\bea \label{FP_Ps}
P(K,N) = \frac{1}{2} P(K,N-1) + \frac{1}{2} P(K-1,N-1) \;, \; \quad N \geq 3 \; {\rm and}\; K\geq 0\;,
\eea
with the convention that $P(-1,N)=0$ for all $N \geq 2$. For $N=2$, direct inspection gives 
\bea \label{PK2}
P(K,2) = \frac{1}{2} \delta_{K,0} + \frac{1}{2} \delta_{K,1} \;.
\eea
We define the double generating function 
\bea \label{GF_PK}
S(u,z) = \sum_{N=2}^\infty \sum_{K=0}^\infty P(K,N)\, u^K z^N \;.
\eea
From Eq. (\ref{FP_Ps}) it is easy to show that $S(u,z)$ is given explicitly by
\bea \label{S_uz_PK}
S(u,z) = \frac{\frac{z^2}{2}(1+u)}{1- \frac{z}{2}(1+u)} \;.
\eea
Expanding in powers of $z$ we get
\bea \label{exp_GF1}
\sum_{K=0}^\infty P(K,N)\,u^K = \frac{1}{2^{N-1}}(1+u)^{N-1} \;.
\eea
Expanding in powers of $u$ immediately gives the binomial result in Eq. (\ref{PkN}). In Fig. \ref{FigPkN}, we compare this theoretical prediction in Eqs. (\ref{PkN}) with numerical simulations for five different jump distributions, showing a perfect agreement.

From this formula (\ref{PkN}) it is easy to calculate all the moments of $K$. For example the mean and the variance are given by
\bea 
\langle K \rangle = \frac{N-1}{2} \quad, \quad {\rm Var}(K) = \langle K^2 \rangle - \langle K \rangle^2 = \frac{N-1}{4} \;. \label{mean_var}
\eea
Since $K = m + M$, we have
\bea \label{VarK}
{\rm Var}(K) = {\rm Var}(m) + {\rm Var}(M) + 2 C_c(N)  \;,
\eea
where $C_c(N) = \langle m M \rangle - \langle m \rangle  \langle M \rangle$ is the connected correlation function between the number of minima $m$ and the number of maxima $M$. Since $M$ has the same statistics as $m$ by symmetry, it follows from Eq. (\ref{VarK}) that 
\bea \label{CcN}
C_c(N) = \frac{{\rm Var}(K) - 2 {\rm Var}(m)}{2} = \frac{N-3}{16} \;,
\eea
where we have used the results for ${\rm Var}(K)$ in Eq. (\ref{mean_var}) and for ${\rm Var}(m)$ in Eq. (\ref{mean_varm}). This result matches perfectly with the result obtained from the joint distribution ${\cal Q}(m,M,N)$ in Eq. (\ref{tildeC}) and verified numerically in Fig. \ref{fig:CMm-n}.

\subsection{Application to the run-and-tumble particle}\label{sec:RTP}

\begin{figure}[t]
\includegraphics[width = 0.5\linewidth]{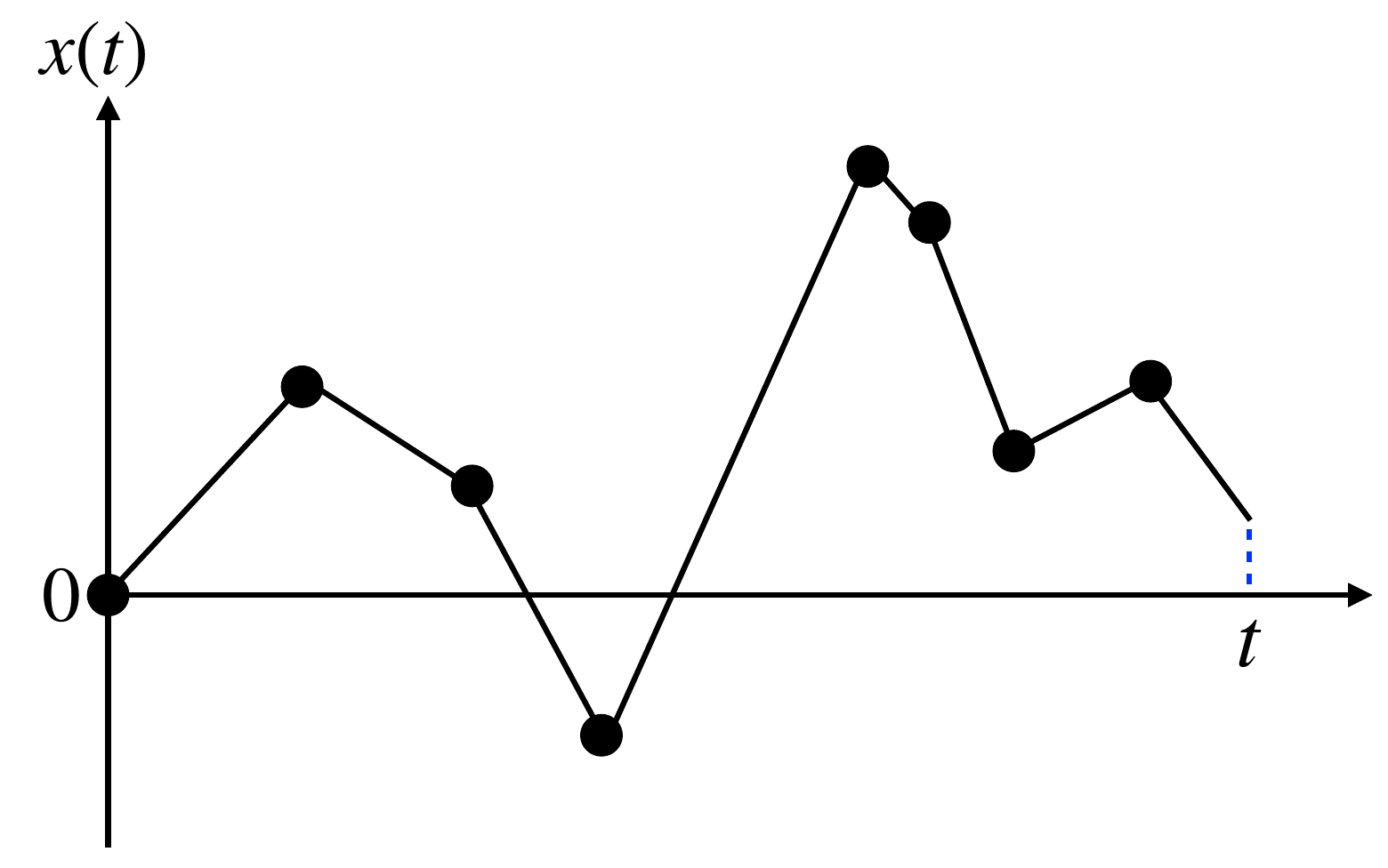}
\caption{A typical trajectory $x(t)$ denoting the $x$-component of an RTP of duration $t$. The straight lines show the $x$-projections of the successive runs and the filled circles denote the tumbles following each run. The last run before $t$ is incomplete.}\label{Fig_RTP}
\end{figure}

In this subsection, we apply our results obtained in the previous subsections for the distribution of the number of stationary points as well as that of the number of minima for a random walk of $N$ steps, to the problem of a run-and-tumble particle (RTP) up to a total duration $t$. Let us recall the definition of the RTP model in $d$ dimensions. A particle, such as an {\it E-Coli} bacteria, starting at the origin, chooses a random velocity ${\bf v}$ drawn from an arbitrary isotropic distribution $W(|{\bf v}|)$ and moves ballistically with this velocity during a random time $\tau$ distributed as $p(\tau) = \gamma\, e^{-\gamma \tau}$. At the end of this run, the particle tumbles, i.e., it chooses a new velocity, again from $W(|{\bf v}|)$ and a new run time $\tau$, from $p(\tau)$. The runs and tumblings alternate. We consider this process up to a final fixed time $t$. Now let us consider the $x$-component of this $d$-dimensional process denoted by $x(t)$ and this projected trajectory of duration $t$ constitutes a one-dimensional landscape, starting at the origin, and consisting of peaks and troughs (see Fig. \ref{Fig_RTP}). These peaks and troughs are the stationary points of this one-dimensional landscape and their number indicates the number of direction reversals that the RTP undergoes in time $t$. We are interested in computing the distribution of the stationary points $P(K,t)$ and also the distribution $Q(m,t)$ of the number of minima (troughs) till time $t$. 

To proceed, we consider the process $x(t)$ representing the $x$-component of the $d$-dimensional RTP trajectory in continuous time as in Fig. \ref{Fig_RTP}. This can be viewed as a discrete-time random walk landscape but
with the number of steps $N(t)$ as a fluctuating random variable, for fixed $t$. Note that the number of steps $N(t)$ is also precisely the number of runs in the RTP trajectory. Our first goal is to compute the distribution $P(N(t) = N \vert t)$ of the number of runs in time $t$. To compute this, 
it is convenient to first introduce the joint distribution $P(\tau_1, \tau_2, \cdots, \tau_N, N(t)=N \vert t)$ of the run times $\{\tau_1, \tau_2, \cdots, \tau_N\}$ and the number $N(t)$ of runs till time $t$, which we assume is fixed. This joint distribution can be written down very simply, since the successive run times are statistically independent, namely~\cite{Mori_SM,Mori_PRE_SM}
\bea \label{joint_tau}
P(\tau_1, \tau_2, \cdots, \tau_N, N(t)=N \vert t) = \left[ \prod_{i=1}^{N-1} p(\tau_i) \right] q(\tau_N) \, \delta\left(\sum_{i=1}^N \tau_i - t\right) \;,
\eea 
where $p(\tau) = \gamma\,e^{-\gamma \tau}$ and $q(\tau) = \int_{\tau}^\infty p(\tau')\,d\tau'$. This can be understood as follows: the first $(N-1)$ runs are complete and each is distributed independently via $p(\tau)$ -- this explains the product in Eq. (\ref{joint_tau}). The last run $\tau_N$ is yet to be complete and hence it is distributed via $q(\tau_N)= \int_{\tau_N}^\infty p(\tau')\,d\tau'$, which comes from the fact that the completion time has to occur after $\tau_N$. Finally the delta function ensures that the total time spent is $t$. Taking the Laplace transform with respect to $t$ and integrating over $\tau_i$'s one finds
\bea \label{P_of_N}
\int_0^\infty P(N \vert t)\, e^{-s t}\, dt = \left[ \tilde p(s)\right]^{N-1} \tilde q(s) \;,
\eea 
where $P(N \vert t) = \int_0^\infty d\tau_1 \cdots \int_0^\infty d\tau_N P(\tau_1, \tau_2, \cdots, \tau_N, N \vert t)$ is the marginal distribution of $N(t)$, given $t$ and $\tilde p(s) = \int_0^\infty p(\tau)\, e^{-s\tau}\, d\tau$ and $\tilde q(s) = \int_0^\infty q(\tau)\, e^{-s\tau}\, d\tau$ are the Laplace transforms of $p(\tau)$ and $q(\tau)$. Using $p(\tau) = \gamma e^{-\gamma \tau}$ and $q(\tau) = e^{-\gamma \tau}$, one gets from Eq. (\ref{P_of_N})
\bea \label{P_of_N.2}
\int_0^\infty P(N \vert t)\, e^{-s t}\, dt = \frac{\gamma^{N-1}}{(\gamma+s)^{N}} \;.
\eea
Inverting this Laplace transform, one gets the Poisson distribution~\cite{Mori_SM,Mori_PRE_SM}
\bea \label{P_of_N.3}
P(N \vert t) = e^{- \gamma t} \frac{(\gamma t)^{N-1}}{(N-1)!} \quad, \quad N=1,2, \cdots \;.
\eea   
 
\begin{figure}[t]
\includegraphics[width= 0.5\linewidth]{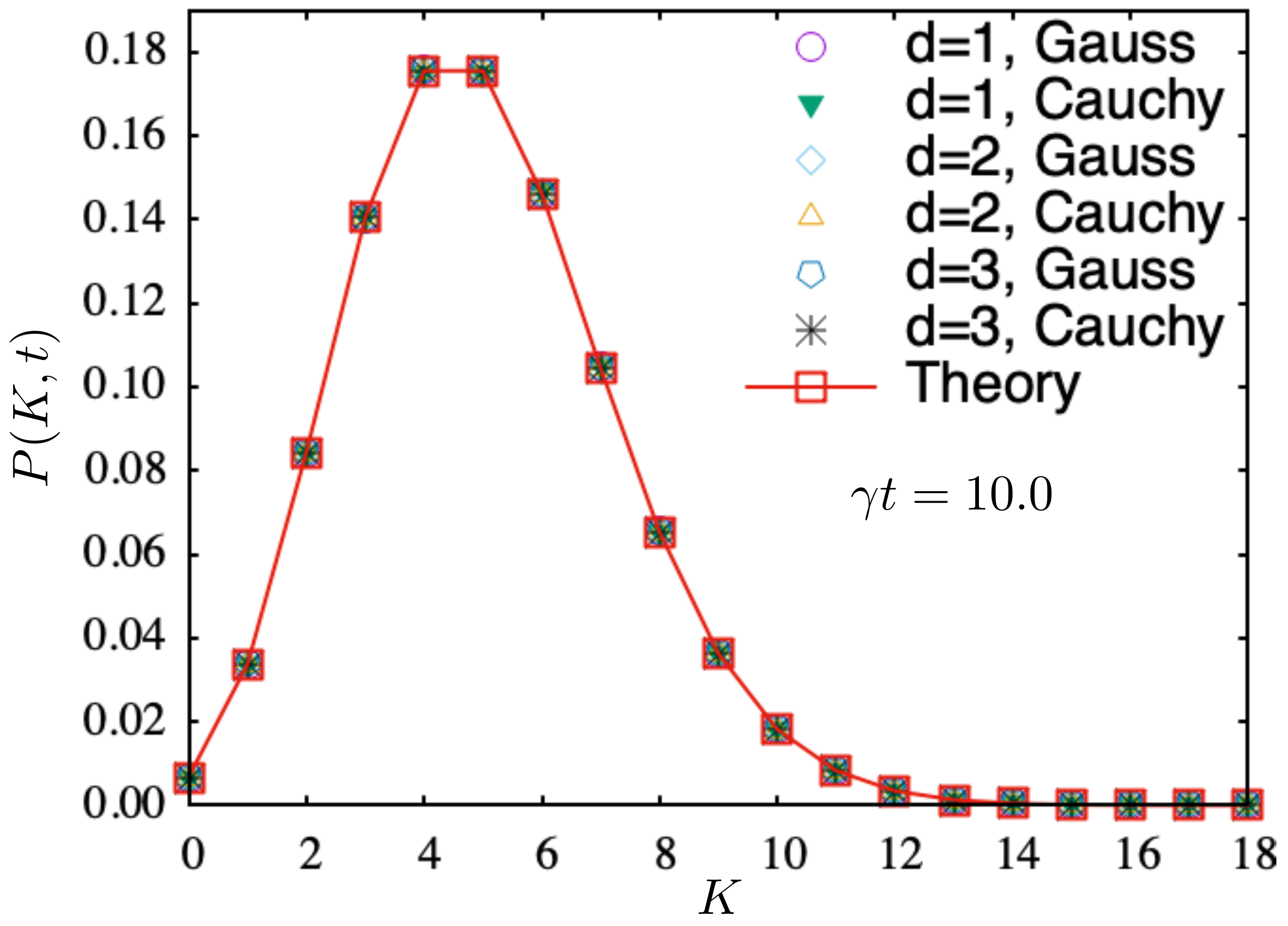}
\caption{Plot of the numerically obtained $P(K,t)$ vs $K$ for an RTP in different dimensions $d=1,2$ and $3$ and for different velocity distributions $W(|{\bf v}|)$ (Gaussian and Cauchy distributions) with $\gamma t = 10$, compared to the theoretical expression, which is a Poissonian distribution with mean $\gamma t/2$ given in Eq.~\eqref{P_K_t}, plotted with square symbols, showing a perfect agreement. The collapse of the data for different dimensions $d$ and different velocity distributions $W(|{\bf v}|)$ on a single curve clearly demonstrates  the universality of $P(K,t)$ for an RTP.} \label{Fig_PKt}
\end{figure}
We now consider the distribution $P(K,t)$ of the total number of stationary points $K$ up to time $t$ in the RTP. We have seen in the previous subsection that 
the distribution $P(K,N)$ of the number of stationary points in a discrete-time random walk of $N$ steps is given exactly as in Eq. (\ref{PkN}). Now for the RTP problem, $N$ itself is a random variable distributed via Eq. (\ref{P_of_N.3}). Taking the product of the two and summing over all $N = 1, 2, \cdots$, we get 
\bea \label{P_K_t}
P(K,t) = \sum_{N=1}^\infty \frac{1}{2^{N-1}} {N-1 \choose K} e^{-\gamma\,t} \frac{(\gamma\,t)^{N-1}}{(N-1)!} = e^{-\frac{\gamma t}{2}} \frac{\left( \frac{\gamma\,t}{2}\right)^{K}}{K!} \quad, \quad {\rm for \; all} \quad K=0,1, \cdots \;.
 \eea
Therefore for fixed $t$, the distribution $P(K,t)$ of the number of stationary points is also a Poisson distribution with mean $\gamma\,t/2$. In Fig. (\ref{Fig_PKt}) we give a plot of $P(K,t)$ vs $K$ for a fixed $\gamma t = 10$ and compare this result with numerical simulations.

\begin{figure}[t]
\includegraphics[width= 0.5\linewidth]{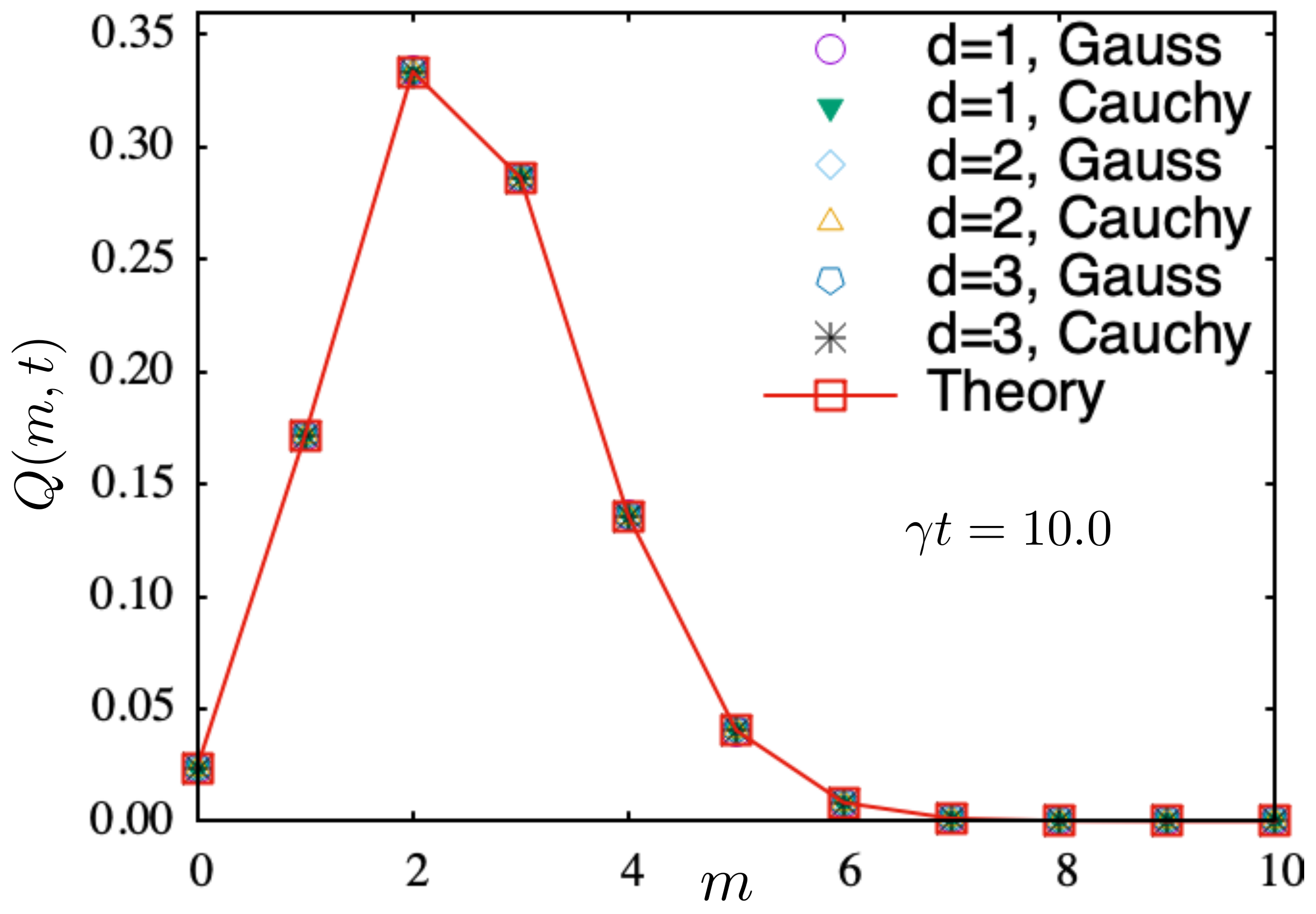}
\caption{Plot of the numerically obtained $Q(m,t)$ vs $m$ for an RTP in different dimensions $d=1,2$ and $3$ and for different velocity distributions $W(|{\bf v}|)$ (Gaussian and Cauchy distributions) with $\gamma t = 10$, compared to the theoretical expression in Eq.~\eqref{Q_m_t.3}, plotted with square symbols, showing a perfect agreement. The collapse of the data for different dimensions $d$ and different velocity distributions $W(|{\bf v}|)$ on a single curve
clearly demonstrates  the universality of $Q(m,t)$ for an RTP.} \label{Fig_Qmt}
\end{figure}

We now turn to the distribution $Q(m,t)$ of the number of minima $m$ up to time $t$ in the RTP trajectory. We have seen in the previous subsections that the distribution $Q(m,N)$ for the number of minima $m$ in a discrete-time random walk trajectory of $N$ steps is given by the exact formula in Eq. (\ref{sol:Q(m,N)}). Using again the fact that, for fixed $t$, the number of runs $N$ itself is a random variable distributed via Eq. (\ref{P_of_N.3}), we get
\bea \label{Q_m_t}
Q(m,t) = \sum_{N=1}^\infty Q(m,N) \,e^{- \gamma t} \frac{(\gamma t)^{N-1}}{(N-1)!} = e^{- \gamma t}\,\sum_{N=1}^\infty \frac{1}{2^N} \frac{(N+1)!}{(2m+1)!(N-2m)!} \frac{(\gamma t)^{N-1}}{(N-1)!} \;.
\eea 
Simplifying and performing a shift $N \to N + 2m$ we get
\bea \label{Q_m_t.2}
Q(m,t) = e^{- \gamma t}\, \frac{\left( \gamma\,t/2\right)^{2m-1}}{(2m+1)!} \; \sum_{N=0}^\infty \frac{(N+2m)(N+2m+1)}{N!} \left( \frac{\gamma \,t}{2}\right)^{N} \;.
\eea
Fortunately, this sum can be performed explicitly (using Mathematica) and we get
\bea \label{Q_m_t.3}
Q(m,t) = e^{- \gamma t/2} \frac{\left( \gamma\,t/2\right)^{2m-1}}{2(2m+1)!} \left[ (2m+1)(2m+\gamma t) + \left( \frac{\gamma \,t}{2}\right)^2 \right] \quad, \quad {\rm for \; all} \quad m=0,1, \cdots \;.
\eea
This is clearly a highly non-Poissonian distribution, unlike the distribution of the number of stationary points in Eq.~(\ref{P_K_t}). For example, the mean and the variance are not equal unlike in the Poisson distribution, but rather are given~by 
\bea \label{cumul_Qm}
\langle m \rangle = \sum_{m=0}^\infty m\, Q(m,t) = \frac{\gamma t}{4} \quad , \quad {\rm Var}(m) = \langle m^2 \rangle - \langle m \rangle^2 = \frac{1}{8}\left(1-e^{-\gamma t}+ \gamma t)\right) \;.
\eea
In Fig. (\ref{Fig_Qmt}) we give a plot of $Q(m,t)$ vs $m$ for a fixed $\gamma t = 10$ and compare this result with numerical simulations. 

Let us remark that the results for $P(K,t)$ in Eq. (\ref{P_K_t}) and for $Q(m,t)$ in Eq. (\ref{Q_m_t.3}) are universal, i.e., independent of the spatial dimension $d$ as well as the velocity distribution $W(|{\bf v}|)$ (see Figs. \ref{Fig_PKt} and \ref{Fig_Qmt}). The spatial dimension $d$ and the velocity distribution $W(|{\bf v}|)$ do affect the jump distribution of the effective random walk consisting of the RTP runs as steps~\cite{Mori_SM,Mori_PRE_SM}. However, since $P(K,N)$ and $Q(m,N)$ are independent of the jump distributions, and since $P(N \vert t)$ is Poissonian it is clear these jump distributions do not enter in the formulae for $P(K,t)$ and $Q(m,t)$. Furthermore, let us also emphasize that these two quantities are universal for all time $t$, and not just for large $t$.

\section{First-passage ensemble}

\begin{figure}[h]
\includegraphics[width = 0.7\linewidth]{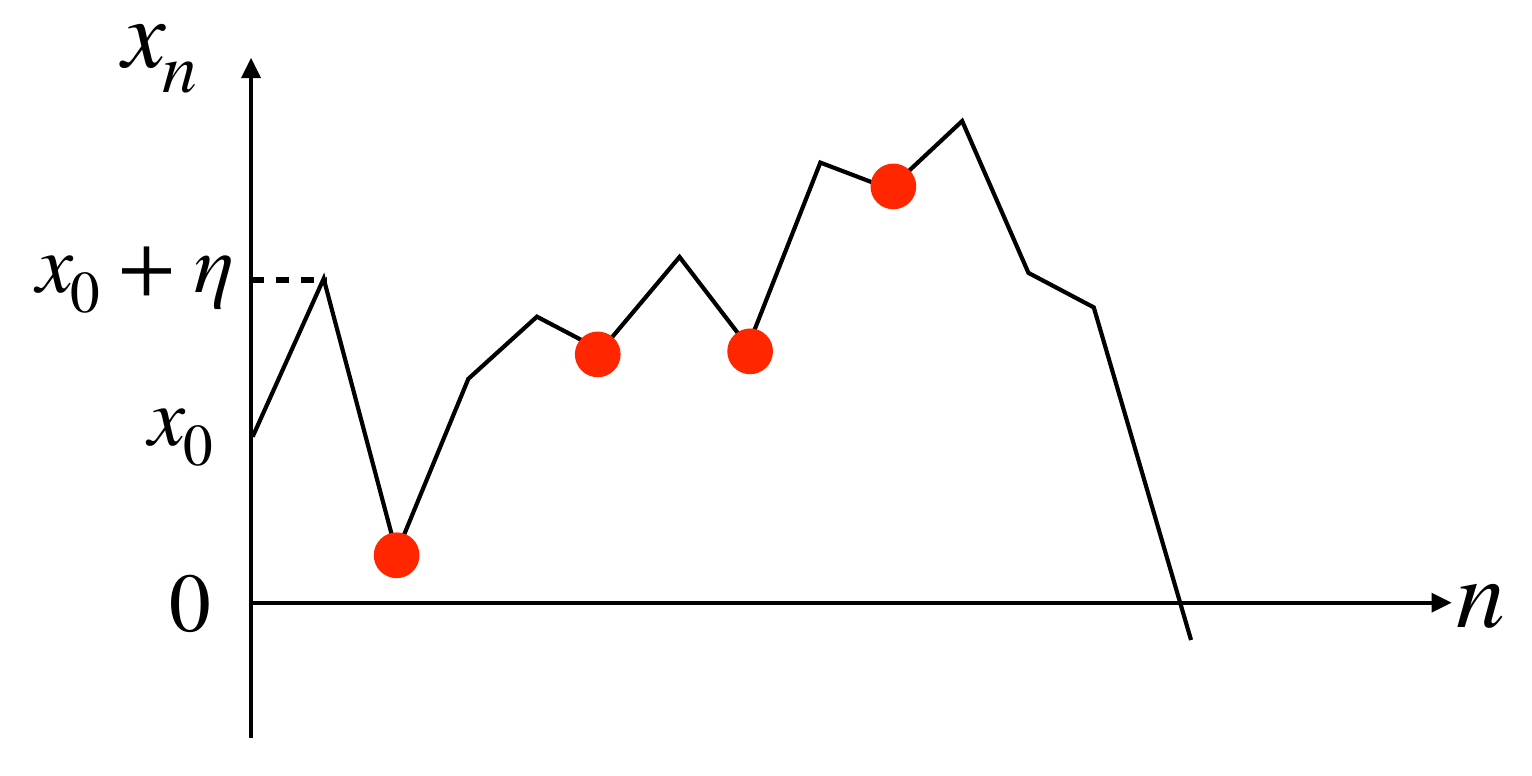}
\caption{A trajectory of a random walk, starting at $x_0 \geq 0$ till the first time it crosses the origin from above. At the first step the walker jumps to $x_0+\eta$ where $\eta$ is a random jump drawn from $\phi(\eta)$ which is symmetric and continuous. The local minima are marked as filled red circles.}\label{Fig_SM_2}
\end{figure}

In this section, we study our random walk in (\ref{def_RW_SM}), starting at initial position $x_0$ and the process stops when the walker crosses the origin for the first time. Without loss of generality, we take $x_0 \geq 0$ (see Fig. \ref{Fig_SM_2}). In the first subsection \ref{section:nb_fp}, we set up the general integral equations for the distribution of the number of minima $Q^{({\rm fp})}(x_0,m)$, starting at $x_0$ and solve explicitly for the special case of a double-exponential jump distribution $\phi(\eta) = (1/2)\, e^{-|\eta|}$. In subsection \ref{section:fp}, we similarly write down the exact integral equations for the distribution $P^{({\rm fp})}(x_0,K)$ of the total number of stationary points $K$, starting at $x_0$, till the first-passage time. Once again, we provide the exact explicit solution for the double-exponential jump distribution $\phi(\eta) = (1/2)\, e^{-|\eta|}$. In subsection \ref{sec:fp_univ}, we prove that both results for $Q^{({\rm fp})}(x_0=0,m)$ and $P^{({\rm fp})}(x_0=0,K)$ are universal, i.e., independent of the jump distribution $\phi(\eta)$ as long as it is symmetric and continuous.

\subsection{Distribution of the number of minima till the first-passage time to the origin}\label{section:nb_fp}

In this subsection, our goal is to compute the distribution of the number of minima $Q^{\rm fp}(x_0,m)$, for a random walk in Eq. (\ref{def_RW_SM}) starting at $x_0 \geq 0 $, till the first time the walk goes on the negative side. As usual, we define the pair of probabilities 
\begin{align}
Q_\pm^{\rm (fp)}(x_0,m)=\text{Prob.}~
\left(\begin{array}{c} 
\text{that the RW has $m$ minima till its first passage to the } \\
\text{origin, starting from $x_0$ with first jump in the $\pm$ direction} 
\end{array} 
\right)\;.
\label{def:Q_pm(x_0,m)}
\end{align}
Once again, investigating what happens after the first step, one can write down the pair of coupled recursion relations 
\begin{align}
Q_+^{\rm (fp)}(x_0,m)&= \int_0^\infty d\eta~ \phi(\eta)~[Q_+^{\rm (fp)}(x_0+\eta,m) + Q_-^{\rm (fp)}(x_0+\eta,m)] \;, \label{eq:Q_+(x_0,m)} \\
Q_-^{\rm (fp)}(x_0,m)&= \int_{-x_0}^0 d\eta~ \phi(\eta)~[Q_+^{\rm (fp)}(x_0+\eta,m-1) + Q_-^{\rm (fp)}(x_0+\eta,m)] + \delta_{m,0} \int_{-\infty}^{-x_0}\phi(\eta)~d\eta \;. \label{eq:Q_-(x_0,m)}
\end{align}
If the first jump is positive and the particle arrives as $x_0+\eta$ with $\eta > 0$, then if the second jump is either positive or negative, no new minimum is
created and integrating over all positive $\eta$ gives Eq. (\ref{eq:Q_+(x_0,m)}). In contrast, if the first jump is negative, there are two possibilities: (i) either
$x_0+\eta >0$ and in this case, if the second jump is positive, a minimum is created while no minimum occurs if the second jump is negative. This explains the first two terms in Eq. (\ref{eq:Q_-(x_0,m)}), and (ii) if $x_0 + \eta < 0$, then the position becomes negative after this first jump and hence the process ends. In this latter case, the number of minima is clearly zero and this explains the last term in Eq. (\ref{eq:Q_-(x_0,m)}).   
These equations are valid for $m \ge 0$ with the interpretation $Q_+(x_0,-1)=0$. To solve these equations we define the following generating functions 
\begin{align}
Z_\pm^{\rm (fp)}(x_0,u)=\sum_{m=0}^\infty Q_\pm^{\rm (fp)}(x_0,m)~u^m, \label{def:Z_pm(x_0,u)}
\end{align}
which, from Eq.~\eqref{eq:Q_+(x_0,m)} and Eq.~\eqref{eq:Q_-(x_0,m)}, can be shown to satisfy the following equations
\begin{align}
Z_+^{\rm (fp)}(x_0,u) &= \int_{x_0}^\infty dy~\phi(y-x_0)~[Z_+^{\rm (fp)}(y,u) + Z_-^{\rm (fp)}(y,u)],   \label{eq:Z_+(x_0,m)} \\
Z_-^{\rm (fp)}(x_0,u) &= \int_{0}^{x_0} dy~\phi(y-x_0)~[uZ_+^{\rm (fp)}(y,u) + Z_-^{\rm (fp)}(y,u)] + \int_{-\infty}^{-x_0}\phi(\eta)~d\eta \;.  \label{eq:Z_-(x_0,m)}
\end{align}
These equations are valid for general jump distributions $\phi(\eta)$. However, they are of the Wiener-Hopf types and are 
hard to solve for arbitrary $\phi(\eta)$. Below, we consider the specific choice of the double-exponential jump distribution $\phi(\eta) = \frac{1}{2}\exp(-|\eta|)$ for which we show that these equations can be solved explicitly. 

\vspace*{0.5cm}
\noindent{\bf Double-exponential jump distribution $\phi(\eta) = \frac{1}{2}\exp(-|\eta|)$.}  For this case Eqs. \eqref{eq:Z_+(x_0,m)} and \eqref{eq:Z_-(x_0,m)} reduce to
\begin{align}
e^{-x_0}Z_+^{\rm (fp)}(x_0,u) &= \frac{1}{2}\int_{x_0}^\infty dy~e^{-y}~[Z_+^{\rm (fp)}(y,u) + Z_-^{\rm (fp)}(y,u)],   \label{eq:Z_+(x_0,m)-1} \\
e^{x_0}Z_-^{\rm (fp)}(x_0,u) &= \frac{1}{2}\int_{0}^{x_0} dy~e^y~[uZ_+^{\rm (fp)}(y,u) + Z_-^{\rm (fp)}(y,u)] + \frac{1}{2}.  \label{eq:Z_-(x_0,m)-1}
\end{align}
Taking derivatives on both sides of the above equations with respect to $x_0$ yields
\begin{align}
e^{-x_0}\left[ \frac{d}{dx_0}-1\right]Z_+^{\rm (fp)}(x_0,u) &= -\frac{1}{2}~e^{-x_0}~\left[Z_+^{\rm (fp)}(x_0,u) + Z_-^{\rm (fp)}(x_0,u)\right],   \label{deq:Z_+(x_0,m)} \\
e^{x_0}\left[ \frac{d}{dx_0}+1\right]Z_-^{\rm (fp)}(x_0,u) &= \frac{1}{2}~e^{x_0}~[uZ_+^{\rm (fp)}(x_0,u) + Z_-^{\rm (fp)}(x_0,u)] \;.  \label{deq:Z_-(x_0,m)}
\end{align}
Simplifying one gets
\begin{align}
\left[ \frac{d}{dx_0}-\frac{1}{2}\right]Z_+^{\rm (fp)}(x_0,u) &= -\frac{1}{2}~ Z_-^{\rm (fp)}(x_0,u) \;,   \label{deq:Z_+(x_0,m)-1} \\
\left[ \frac{d}{dx_0}+\frac{1}{2}\right]Z_-^{\rm (fp)}(x_0,u) &= \frac{1}{2}~uZ_+^{\rm (fp)}(x_0,u) \;.   \label{deq:Z_-(x_0,m)-1}
\end{align}
It is easy to see that these two equations can be re-written as 
\begin{align}
\left[ \frac{d^2}{dx_0^2} - \frac{1}{4}\right] Z_\pm^{\rm (fp)}(x_0,u) = - \frac{u}{4}Z^{\rm fp}_\pm(x_0,u).  \label{deq:Z_pm(x_0,m)} 
\end{align}
It is clear from the definition of the generating functions $Z_\pm(x_0,u)$ in Eq.~\eqref{def:Z_pm(x_0,u)} that they can not diverge exponentially as the starting point $x_0 \to \infty$. Using this condition, we get the two solutions 
\begin{align}
Z_+^{\rm (fp)}(x_0,u) &=A~\exp\left( -\frac{\sqrt{1-u}}{2}x_0\right), \label{sol:Z_+} \\
Z_-^{\rm (fp)}(x_0,u) &=B~\exp\left( -\frac{\sqrt{1-u}}{2}x_0\right) \;, \label{sol:Z_-} 
\end{align}
where the two constants $A$ and $B$ are yet to be determined. Setting $x_0=0$ in Eq. (\ref{eq:Z_-(x_0,m)-1}) one immediately gets $Z_-^{\rm fp}(0,u) = 1/2$. Using this condition in Eq. (\ref{sol:Z_-}) it follows that
\bea \label{eq_B}
B = \frac{1}{2} \;.
\eea
To fix the other constant $A$, we proceed as follows. The integral equations (\ref{eq:Z_+(x_0,m)-1}) and (\ref{eq:Z_-(x_0,m)-1}) actually contain more informations than the derived differential equations (\ref{deq:Z_+(x_0,m)-1}) and (\ref{deq:Z_-(x_0,m)-1}). Hence one has to additionally ensure that the solutions of the differential equation also satisfy the integral equations. Indeed, substituting Eqs. (\ref{sol:Z_+}) and (\ref{sol:Z_-}) in the integral equations, one sees that these are indeed the solutions provided 
\begin{align}
A&=\frac{1-\sqrt{1-u}}{2u}, ~~
B=\frac{1}{2} \;. \label{sol:AB}
\end{align}
Hence we have the following explicit solution of $Z^{\rm (fp)}_\pm(x_0,u)$:
\begin{align}
Z_+^{\rm (fp)}(x_0,u) &=\frac{1-\sqrt{1-u}}{2u}~\exp\left( -\frac{\sqrt{1-u}}{2}x_0\right), \label{sol:Z_+-ex} \\
Z_-^{\rm (fp)}(x_0,u) &=\frac{1}{2}~\exp\left( -\frac{\sqrt{1-u}}{2}x_0\right). \label{sol:Z_--ex} 
\end{align}
Now to find $Q_\pm(x_0,m)$ one requires to perform the inverse transform given by the Cauchy formula
\begin{align}
Q_\pm^{\rm (fp)}(x_0,m) = \frac{1}{2\pi i} \oint du~\frac{1}{u^{m+1}}~Z^{\rm(fp)}_\pm(x_0,u). \label{cauchy-formula}
\end{align}

\begin{figure}
\centering
    \includegraphics[width=0.5\linewidth]{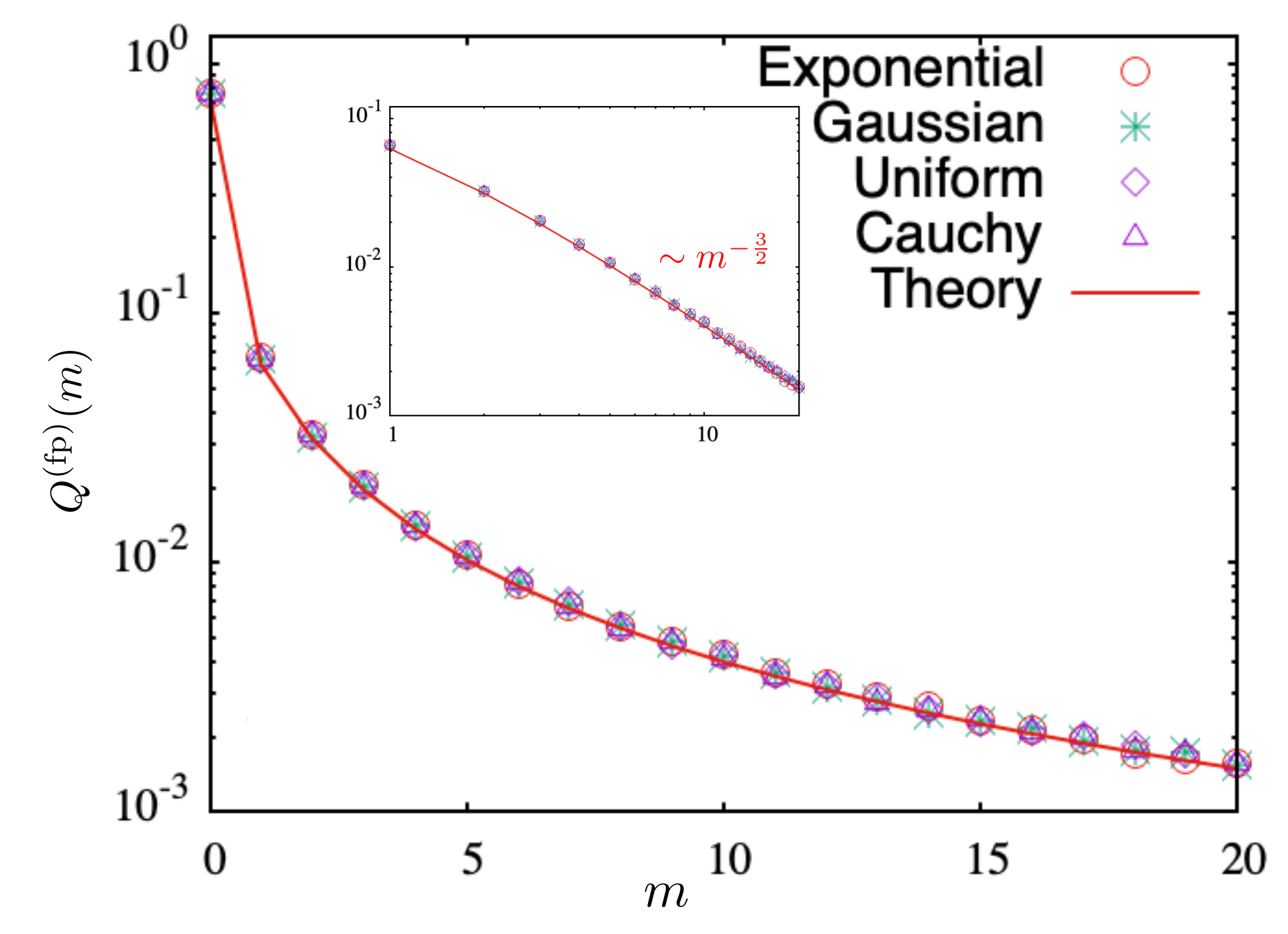}
\caption{Numerical verification of the analytical expression of $Q^{\rm (fp)}(m)=Q^{\rm (fp)}(0,m)$ in Eq.~\eqref{sol:Q(0,m)-fr} obtained for the exponential jump distribution. We have also plotted the distribution $Q^{\rm (fp)}(m)$ obtained (numerically)  for other continuous and symmetric jump distributions $\phi(\eta)$. We find that for other jump distributions also the distribution $Q^{\rm (fp)}(m)$ is given by Eq.~\eqref{sol:Q(0,m)-fr} as long as $\phi(\eta)$ is continuous and symmetric, {\it i.e.}, $\phi(\eta) =\phi(-\eta)$. The simulation results  for different jump distributions
strongly indicate that the result in Eq. (\ref{sol:Q(0,m)-fr}) is universal, which indeed we prove rigorously also. {\bf Inset:} Numerical verification of the $m^{-3/2}$ decay of $Q^{\rm (fp)}(m)$ for large $m$ as predicted in Eq.~\eqref{Q(0,m)-asymp}.}
\label{QmN-gauss-large-N}
\end{figure}

The results become more explicit for the case $x_0=0$. In this case, the expressions of $Z^{\rm(fp)}_\pm(0,u)$ simplifies and for $Z^{\rm (fp)}(0,u)=Z^{\rm (fp)}_+(0,u)+Z^{\rm (fp)}_-(0,u)$ one has 
\begin{align}
Z^{\rm (fp)}(0,u) &= \frac{1}{2}+\frac{1}{2u}\left(1-\sqrt{1-u} \right), \notag \\ 
&= \frac{3}{4} +\frac{1}{2} ~\sum_{k=1}^\infty\frac{1}{2^{2k+1}}~\frac{(2k)!}{k!(k+1)!}~u^k. \label{sol:Z(0,u)}
\end{align}
From the coefficient of $u^m$ it is easy to extract 
\begin{align}
Q^{\rm (fp)}(m) =Q^{\rm (fp)}(0,m) = 
\begin{cases}
\frac{3}{4} ~& ~\text{for}~m=0, \\
\frac{1}{2^{2m+2}}~\frac{(2m)!}{m!(m+1)!}~&~\text{for}~m\ge 1.
\end{cases}
\label{sol:Q(0,m)-fr}
\end{align}
This analytical expression is numerically verified in Fig.~\ref{QmN-gauss-large-N} where the solid red line corresponds to the expression in Eq.~\eqref{sol:Q(0,m)-fr} whereas the cross symbols are obtained from simulation for the double-exponential jump distribution $\phi(\eta)=\frac{1}{2}\exp(-|\eta|)$.

For large $m$, the distribution $Q^{\rm (fp)}(m)$ in Eq. (\ref{sol:Q(0,m)-fr}) has a power law tail 
\begin{align}
Q^{\rm (fp)}(m) \overset{m \to \infty}{\approx} \frac{1}{4 \sqrt{2 \pi}}~\frac{1}{m^{3/2}} \;. \label{Q(0,m)-asymp}
\end{align}
This power-law tail can be understood from the following scaling argument. In an $N$-step random walk, the number of minima $m$ typically scales as $m \sim N$ for large $N$. On the other hand, the number of steps till the first-passage time has a power-law distribution $N^{-3/2}$ large $N$, which follows from the Sparre Andersen theorem \cite{SA}. This shows that distribution of $m$ will have the power-law decay with the same exponent $3/2$. Note that while this scaling argument predicts correctly the universal exponent $3/2$, it can not be used to obtain the exact universal prefactor ${1}/{(4 \sqrt{2 \pi})}$ in Eq. (\ref{Q(0,m)-asymp}). 
Below we will prove that, actually, this expression for $Q^{\rm (fp)}(m)$ in Eq. (\ref{sol:Q(0,m)-fr}), derived above only for the double-exponential jump distribution, is actually universal for all $m$, i.e., independent of the jump distribution $\phi(\eta)$, as long as it is symmetric and continuous.

\subsection{Distribution of the number of stationary points till the first-passage time to the origin}\label{section:fp}

In this subsection, we first set up the integral equations for the distribution $P_\pm^{\rm (fp)}(x_0,K)$ denoting the distribution of the number of stationary points (minima and maxima) till the first-passage time to the origin, starting from $x_0$ with either a positive or a negative jump. Examining the different possibilities after the first jump, we can write down the exact recursion relations
\begin{align}
P_+^{\rm (fp)}(x_0,K)&= \int_0^\infty d\eta~ \phi(\eta)~[P_+^{\rm (fp)}(x_0+\eta,K) + P_-^{\rm (fp)}(x_0+\eta,K-1)] \;, \label{eq:P_+(x_0,m)} \\
P_-^{\rm (fp)}(x_0,K)&= \int_{-x_0}^0 d\eta~ \phi(\eta)~[P_+^{\rm (fp)}(x_0+\eta,K-1) + P_-^{\rm (fp)}(x_0+\eta,K)] + \delta_{K,0} \int_{-\infty}^{-x_0}\phi(\eta)~d\eta \;. \label{eq:P_-(x_0,m)}
\end{align}
If the first jump is positive and the particle arrives as $x_0+\eta$ with $\eta > 0$, then if the second jump is also positive then no new stationary point is created. This explains the first term in Eq. (\ref{eq:P_+(x_0,m)}). In contrast, if the second jump is negative, a maximum is created and hence the rest of the trajectory, starting at $x_0+\eta$, must have $K-1$ stationary points, explaining the second term in Eq. (\ref{eq:P_+(x_0,m)}). Similarly, if the first step is negative, such that $x_0+\eta > 0$, then, depending on the second step, we will either have either $K-1$ or $K$ stationary points, starting at $x_0+\eta$. This explains the non-delta function term in Eq. (\ref{eq:P_-(x_0,m)}). On the other hand, if the first jump puts the walker on the negative side, the process stops and we just have zero stationary point, explaining the Kronecker delta-function term in (\ref{eq:P_-(x_0,m)}).    

We define the generating functions 
\begin{align}
\tilde{Z}_\pm^{\rm (fp)}(x_0,u)=\sum_{K=0}^\infty P_\pm^{\rm (fp)}(x_0,K)~u^K, \label{def:tildeZ_pm(x_0,u)}
\end{align}
which, from Eqs.~\eqref{eq:P_+(x_0,m)} and~\eqref{eq:P_-(x_0,m)}, satisfy the following equations
\begin{align}
\tilde Z_+^{\rm (fp)}(x_0,u) &= \int_{x_0}^\infty dy~\phi(y-x_0)~\left[\tilde Z_+^{\rm (fp)}(y,u) + u\,\tilde Z_-^{\rm (fp)}(y,u)\right],   \label{eq:tildeZ_+(x_0,m)} \\
\tilde Z_-^{\rm (fp)}(x_0,u) &= \int_{0}^{x_0} dy~\phi(y-x_0)~\left[u\,\tilde Z_+^{\rm (fp)}(y,u) + \tilde Z_-^{\rm (fp)}(y,u)\right] + \int_{-\infty}^{-x_0}\phi(\eta)~d\eta \;.  \label{eq:tildeZ_-(x_0,m)}
\end{align}

\vspace*{0.5cm}
\noindent{\bf Double-exponential jump distribution $\phi(\eta) = \frac{1}{2}\exp(-|\eta|)$.}  For this case Eqs. \eqref{eq:tildeZ_+(x_0,m)} and \eqref{eq:tildeZ_-(x_0,m)} reduce to
\begin{align}
e^{-x_0}\tilde Z_+^{\rm (fp)}(x_0,u) &= \frac{1}{2}\int_{x_0}^\infty dy~e^{-y}~[\tilde Z_+^{\rm (fp)}(y,u) + u\,\tilde Z_-^{\rm (fp)}(y,u)],   \label{eq:tildeZ_+(x_0,m)-1} \\
e^{x_0}\tilde Z_-^{\rm (fp)}(x_0,u) &= \frac{1}{2}\int_{0}^{x_0} dy~e^y~[u \tilde Z_+^{\rm (fp)}(y,u) + \tilde Z_-^{\rm (fp)}(y,u)] + \frac{1}{2}.  \label{eq:tildeZ_-(x_0,m)-1}
\end{align}
These equations can be solved as in the previous subsection leading to the final result
\begin{align}
\tilde Z_+^{\rm (fp)}(x_0,u) &=\frac{1-\sqrt{1-u^2}}{2u}~\exp\left( -\frac{\sqrt{1-u^2}}{2}x_0\right), \label{sol:Z_+-ex} \\
\tilde Z_-^{\rm (fp)}(x_0,u) &=\frac{1}{2}~\exp\left( -\frac{\sqrt{1-u^2}}{2}x_0\right). \label{sol:Z_--ex} 
\end{align}

The result becomes more explicit for the case $x_0=0$. In this case, we find that $\tilde Z^{\rm (fp)}(0,u)=\tilde Z^{\rm (fp)}_+(0,u)+\tilde Z^{\rm (fp)}_-(0,u)$ is simply 
\begin{align}
\tilde Z^{\rm (fp)}(0,u) &= \sum_{K=0}^\infty P^{({\rm fp})}(0,K)\,u^K = \frac{1}{2} + \frac{1}{2u} \left(1 - \sqrt{1-u^2} \right) \;. \label{sol:tildeZ(0,u)}
\end{align}
Expanding in powers of $u$, we get the final result
\begin{eqnarray}
P^{({\rm fp})}(0,K=0) &=& \frac{1}{2} \label{sol:P(0,m)-fr0} \\
 P^{({\rm fp})}(0,K=2\ell-1) &=& \frac{1}{2} (q_{\ell-1}-q_\ell) \;, \; {\rm for}\; \ell \geq 1 \;, \label{sol:P(0,m)-fr}
\end{eqnarray}
where $q_{\ell}$ is given by
\bea \label{q_ell}
q_{\ell} = \frac{1}{2^{2\ell}} {2\ell \choose \ell} \;.
\eea
Note that the number of stationary points $K$ is either $0$ or an odd number. It can not be an even positive number, as evident from examining any typical first-passage trajectory (see for example Fig. \ref{Fig_SM_2} which has $K = 9$ stationary points).

\subsection{The general universal results for $Q^{({\rm fp})}(x_0=0,m)$ and $P^{({\rm fp})}(x_0=0,K)$ for generic continuous and symmetric jump distribution $\phi(\eta)$}\label{sec:fp_univ}

We have seen that, for general $\phi(\eta)$, the distributions $Q_\pm^{({\rm fp})}(x_0,m)$ (for the number of minima) and $P_\pm^{({\rm fp})}(x_0,K)$ (for the number of stationary points) satisfy respectively the coupled integral equations (\ref{eq:Q_+(x_0,m)})-(\ref{eq:Q_-(x_0,m)}) and (\ref{eq:P_+(x_0,m)})-(\ref{eq:P_-(x_0,m)}). These integral equations are hard to solve for general $\phi(\eta)$. However, it turns out that at the special point $x_0=0$, the results for $Q_\pm^{({\rm fp})}(x_0=0,m)$ and $P_\pm^{({\rm fp})}(x_0=0,K)$ are universal, i.e., independent of $\phi(\eta)$. This universal result is hard to prove from the integral equation. However, we found an alternative method via an exact mapping to an auxiliary random walk that allows us to prove this universal result. We present this proof below, starting with $Q_\pm^{({\rm fp})}(x_0=0,m)$ and followed by $P_\pm^{({\rm fp})}(x_0=0,K)$. 
It turns out to be easier to consider the cumulative probability of the number of minima, i.e.,
the probability that the walk, till its first-passage, has {\it at least} $m$ local minima.

\begin{figure}
\centering
\includegraphics[width = 0.7\linewidth]{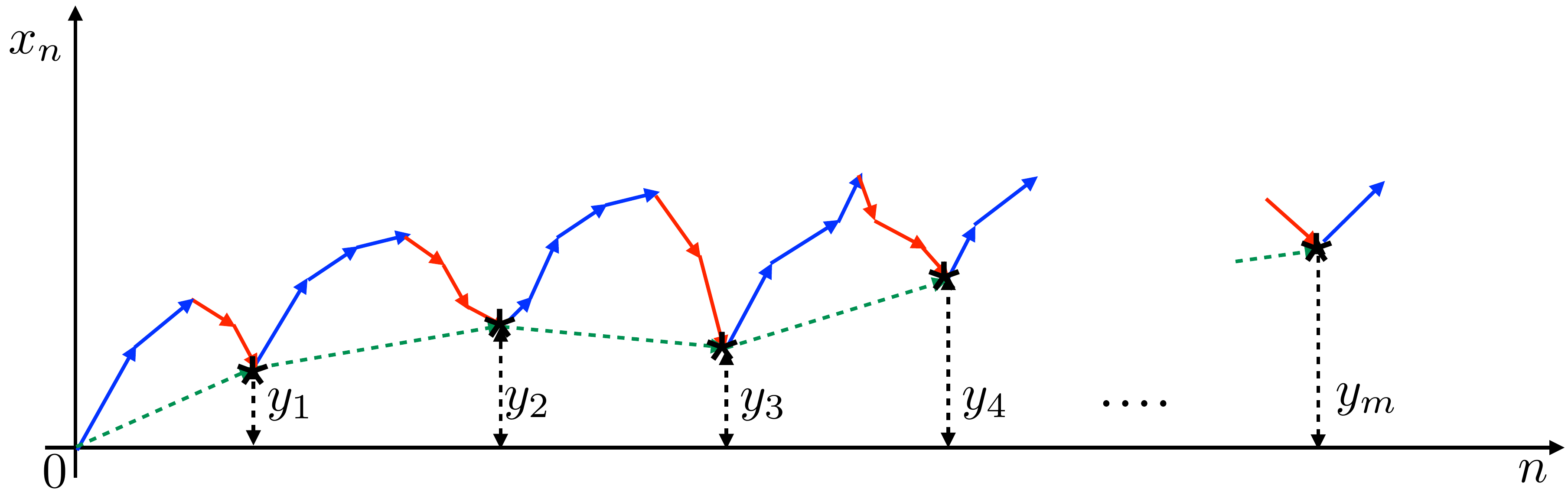}
\caption{Schematic trajectory of a random walk (discrete time and
continuous space) that starts at the origin $x_0=0$, stays
non-negative and has {\it  at least} $m$ local minima. The $*$'s denote the local minima of this configuration
(not counting the starting position $0$) and 
$\{y_1\ge 0,\,  y_2\ge 0,\,  y_3\ge 0\, \ldots\, ,\,  y_m\ge 0 \}$ denote
 the heights of the successive local minimum. The configuration till the $m$-th minimum can then be 
broken into $m$ blocks or segments separated by the dashed vertical lines.
Each block contains only one peak (maxima). One can construct an effective
auxiliary random walk that jumps from one minimum to the next minimum (of the original RW) 
with positions 
$\{ y_1\ge 0,\,  y_2\ge 0,\, \ldots\, y_m\ge 0 \}$ 
(as shown by the green dashed
lines). Thus the step number of the auxiliary walk is identified
with the label of the block. The number of steps of the auxiliary walk
is exactly equal to the number of blocks in the configuration, i.e.,
the number of local minima.}
\label{Fig_rw_block}
\end{figure}

To proceed, let us consider a trajectory of the random walk that has not yet crossed the origin, i.e.,
still surviving (see Fig. \ref{Fig_rw_block}) and has {\it  at least} $m$ local minima. 
We first locate the local minima in this configuration
and denote them by $*$'s in Fig. \ref{Fig_rw_block}. Let $\{y_1\ge 0,\, y_2\ge 0,\, y_3\ge 0,\ldots\, ,\, y_m\ge 0\}$ denote the heights (position) of the 
successive local minimum.
In this configuration $y_i\ge 0$ for all $i=1,2,\ldots, m$, since the walk is surviving. Thus the configuration
till the $m$-th minimum can be broken into $m$ blocks or segments separated by the dashed vertical lines showing the times
of occurrences of the local minima. By construction, each block contains only one peak (in the
interior of the block and not at its edges). If we can integrate out all one-peak configurations
between two successive local mimima, we can then construct a new `auxiliary' or effective random walk
that jumps from one minimum to the next in `one' step (shown by the green dashed lines in
Fig. \ref{Fig_rw_block}) with positions $\{y_1,\,y_2,\,y_3\ldots\}$. The transition probability density of
this walk is just the probability $G(y_1,y_2)$ that the random walk, starting at $y_1$ will
arrive at $y_2$ (in arbitrary number of original steps) with only one peak (or turnaround) in between.
After that we have to ensure that this auxiliary walk stays positive, i.e., $y_1\ge 0$, $y_2\ge 0$ etc. Below we first
compute this transition probability $G(y_1,y_2)$ for general $y_1$ and $y_2$ (not necesarily positive) 
of this auxiliary walk. Then we will come back to the question
of the survival probability of this auxiliary walk.

\vskip 0.3cm

\begin{figure}[t]
\centering
\includegraphics[width = 0.5\linewidth]{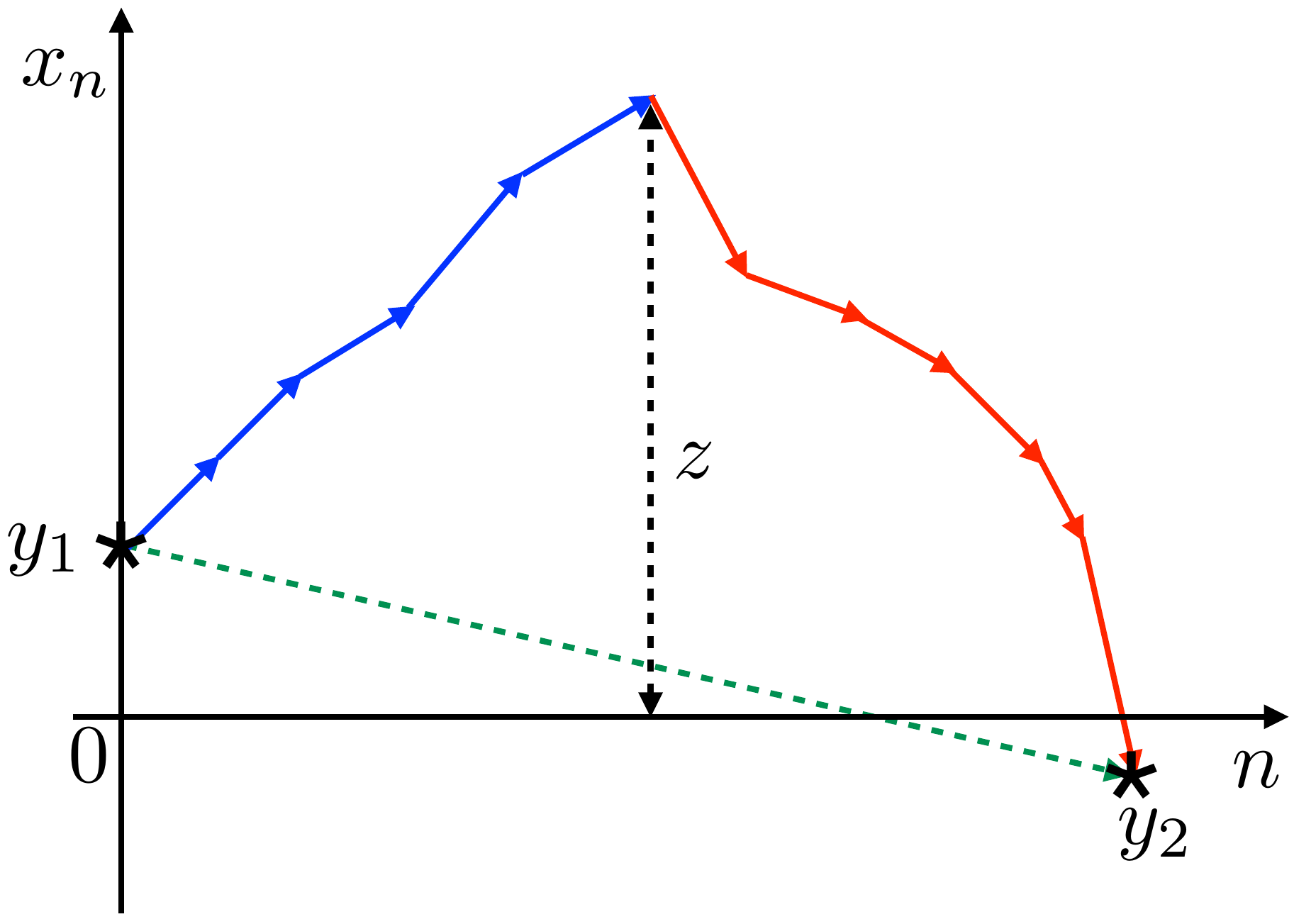}
\caption{Schematic trajectory of a random walk (discrete time and
continuous space), that starts at $y_1$ and arrives at $y_2$ (in arbitrary number of steps),
but with the constraint that there is only one peak (maximum) in the interior (and not
at the edge). This constraint allows only configurations where the walker,
starting at $y_1$ moves up to the peak height $z\ge y_1$ by consecutive upward steps
and then from the peak comes down to $y_2\le z$ by consecutive downward moves. One then
integrates over all possible heights $z$ of the peak. The
probability $G(y_1,y_2)$ of this event will provide the jump probability from $y_1$ to $y_2$
for the auxiliary walk (as indicated by the dashed green arrow).} 
\label{Fig_block1}
\end{figure}

\noindent{\bf The transition probability of the auxiliary walk:} We would like to then compute
the probability of transition from $y_1$ to $y_2$ of the original random walk in arbitrary number of
steps for general $y_1$ and $y_2$, not necessarily positive and with
the constraint that there can be only one peak (maximum) in between. A typical configuration is
shown in Fig. (\ref{Fig_block1}). Let $z$ denote the height of this single peak. Since
$z$ is the maximum, we must have $z\ge \max(y_1,y_2)$. Thus the walker,
starting at $y_1$ moves up to the peak height $z$ by consecutive upward steps
and then from the peak comes down to $y_2$ by consecutive downward moves. One then
integrates over all possible heights $z$ of the peak to compute the transition probability
$G(y_1,y_2)$ of the auxiliary walk that jumps from $y_1$ to $y_2$. Let the probability
that the walker, starting at $y_1$, reaches the peak of height $z\ge y_1$ only by consecutive upward moves
be denoted by $\mathscr{P}_{\rm left}(z-y_1)$. This can be easily computed in terms of the jump distribution $\phi(\eta)$ as
follows. 

Suppose $\mathscr{P}_k(x)$ denotes the probability that the walk takes $k$ consecutive upward steps to reach a point $x$, starting from $0$. 
Then, $\mathscr{P}_{\rm left}(x)$ is simply given by  
\begin{equation}
\mathscr{P}_{\rm left}(x)= \sum_{k=1}^{\infty} \mathscr{P}_k(x)\, .
\label{P_left.1}
\end{equation}
Clearly, 
$\mathscr{P}_k(x)$ satisfies the recursion relation, for $k\ge 1$ and $x\ge 0$,
\begin{equation}
\mathscr{P}_k(x)= \int_0^x dx'\, \mathscr{P}_{k-1}(x')\, \phi(x-x')\, ,
\label{pm_recur.1}
\end{equation}
starting from $\mathscr{P}_0(x)=\delta(x)$. Since $x\ge 0$ and Eq. (\ref{pm_recur.1})
has a convolution form, it is useful to define the Laplace transform
\begin{equation}
\tilde{\mathscr{P}}_k(\lambda)= \int_0^{\infty} \mathscr{P}_k(x)\, e^{-\lambda\, x}\, dx\, .
\label{pm_lap.1}
\end{equation}
Taking Laplace transform of Eq. (\ref{pm_recur.1}) and iterating (using
the initial condition), one gets for $k\ge 1$
\begin{equation}
\tilde{\mathscr{P}}_k(\lambda)= \left[\tilde{\phi}(\lambda)\right]^k \, 
\label{pm_lap.2}
\end{equation}
where
\begin{equation}
\tilde{\phi}(\lambda)= \int_0^{\infty} \phi(\eta)\, 
e^{-\lambda\, \eta}\, d\eta\, .
\label{phi_lap.1}
\end{equation}
Note that
\begin{equation}
\tilde{\phi}(0) = \int_0^{\infty} \phi(\eta)\, d\eta= \frac{1}{2}\, ,
\label{phi_norm.1}
\end{equation}
where we used the symmetry of $\phi(\eta)$. Finally, the Laplace transform
of ${\mathscr{P}}_{\rm left}(x)$ is then given by
\begin{equation}
\tilde{\mathscr{P}}_{\rm left}(\lambda)= \int_0^{\infty}\mathscr{P}_{\rm left}(x)\, e^{-\lambda\, x}\, dx= 
\sum_{k=1}^{\infty} \left[\tilde{\phi}(\lambda)\right]^k
= \frac{\tilde{\phi}(\lambda)}{1- \tilde{\phi}(\lambda)}\, .
\label{P_left.2}
\end{equation} 
The distribution $\mathscr{P}_{\rm left}(x)$ has a support over $x\ge 0$. Furthermore, using Eq. (\ref{phi_norm.1}), 
it follows from (\ref{P_left.2}) that $\tilde{\mathscr{P}}_{\rm left}(\lambda = 0)=1$, indicating that the distribution $\mathscr{P}_{\rm left}(x)$ is normalized to unity, 
\bea \label{norm_Pleft}
\int_0^\infty \mathscr{P}_{\rm left}(x)\, dx = 1 \;.
\eea 
Thus $\mathscr{P}_{\rm left}(x)$ clearly depends explicitly on the jump distribution $\phi(\eta)$. However, we will see below that the detailed form of  
$\mathscr{P}_{\rm left}(x)$ does not really matter for establishing the proof of universality. The only thing that matters is that $\mathscr{P}_{\rm left}(x)$ has a support over $x \geq 0$ and is normalized to unity. 

Getting back to Fig. \ref{Fig_block1}, the probability of reaching $y_2\le z$, starting at $z$, is given by
$\mathscr{P}_{\rm left}(z-y_2)$, where we have simply reversed the steps in order to relate this probability
to the function $\mathscr{P}_{\rm left}(x)$. Finally, integrating over $z\ge {\rm max}(y_1,y_2)$ we get the transition probability
\begin{equation}
G(y_1,y_2)= \int_{\max(y_1,y_2)}^{\infty} dz\, \mathscr{P}_{\rm left}(z-y_1) \, \mathscr{P}_{\rm left}(z-y_2)\, .
\label{G1.1}
\end{equation}
When $y_1>y_2$, the lower limit in the integral is $y_1$ and then making the shift $u=z-y_1$ one gets
\begin{equation}
G(y_1,y_2)= \int_{0}^{\infty} du\, \mathscr{P}_{\rm left}(u) \, \mathscr{P}_{\rm left}(u+y_1-y_2)\, .
\label{G1.2}
\end{equation}
Conversely, when $y_2>y_1$, one similarly gets
\begin{equation}
G(y_1,y_2)= \int_{0}^{\infty} du\, \mathscr{P}_{\rm left}(u) \, \mathscr{P}_{\rm left}(u+y_2-y_1)\, .
\label{G1.3}
\end{equation}
Consequently, the transition probability $G(y_1,y_2)$ depends only on the difference $y_2-y_1$ and is
given by
\begin{equation}
G(y_1,y_2) \equiv \psi(y_2-y_1)= \int_{0}^{\infty} du\, \mathscr{P}_{\rm left}(u) \, \mathscr{P}_{\rm left}(u+|y_2-y_1|)\, .
\label{G_symm}
\end{equation}

Thus, the efective transition probability $\psi(\xi)$ is continuous and symmetric around $\xi=0$. To see that
it is also normalized to unity, we consider the integral
\begin{equation}
\int_{-\infty}^{\infty} \psi(\xi)\, d\xi = \int_{-\infty}^{\infty} d \xi\, \int_0^{\infty} du\, \mathscr{P}_{\rm left}(u)\, 
\mathscr{P}_{\rm left}(u+|\xi|)= \int_0^{\infty} du\, \mathscr{P}_{\rm left}(u)\, \int_{-\infty}^{\infty} d\xi\, \mathscr{P}_{\rm left}(u+|\xi|)\, .
\label{psi_norm.1}
\end{equation}
Next we write
\begin{equation}
\int_{-\infty}^{\infty} d\xi\, \mathscr{P}_{\rm left}(u+|\xi|)= \int_{-\infty}^0 d\xi\,  \mathscr{P}_{\rm left}(u-\xi) + \int_0^{\infty} d\xi\,
 \mathscr{P}_{\rm left}(u+\xi)=2\, \int_u^{\infty} dv\, \mathscr{P}_{\rm left}(v)\, ,
\label{manip.1}
\end{equation}
where we made the change of variable $u-\xi=v$ in the first integral and $u+\xi=v$ in the second one.
Substituting (\ref{manip.1}) in Eq. (\ref{psi_norm.1}) gives
\begin{equation}
\int_{-\infty}^{\infty} \psi(\xi)\, d\xi= 2\, \int_0^{\infty} du\, \mathscr{P}_{\rm left}(u)\, \int_u^{\infty} dv\, \mathscr{P}_{\rm left}(v)= 1\, .
\label{psi_norm.2}
\end{equation}
The last equality is established by making the change of variable $z = \int_u^\infty dv\, \mathscr{P}_{\rm left}(v)$. One sees immediately that
the detailed form of $\mathscr{P}_{\rm left}(x)$ is not important in establishing the normalization of the transition probability $\psi(\xi)$.

Thus in summary, the transition probability $\psi(\xi)$, which is symmetric, continuous and normalized to unity,
can be viewed as an `effective' jump probability associated with the auxiliary random walk process (see Fig. \ref{Fig_rw_block})
\begin{equation}
y_k= y_{k-1} + \xi_k
\label{y_lange.1}
\end{equation}
starting at $y_0=0$, 
where $\xi_k$'s are i.i.d jump variables each drawn from the symmetric and continuous jump density
$\psi(\xi)$ defined in Eq. (\ref{G_symm}). But we will see below that we do not need the explicit form
of the jump density $\psi(\xi)$ for the quantity of our interest.

\vskip 0.3cm

Having computed the jump probability $\psi(\xi)$ of the auxiliary random walk $y_k$ in Eq. (\ref{y_lange.1}),
we now come back to the original question in Fig. \ref{Fig_rw_block}, namely, what is the probability
that the surviving walk has at least $m$ local minima? Using the mapping to the auxiliary process, this is 
equivalent to saying that the auxilary process $y_k$, starting at $y_0=0$, stays positive up
to step $m$ (because to have at least $m$ minima we must have $m$ blocks, i.e., $m$ steps for
the auxiliary walk). But it is not just enough to have $m$ blocks, we also have to ensure that
the last position $y_m$ at the end of the $m$-th block must be a local minimum. In other words,
the step of the original walk immediately following $y_m$ must be upward (only then $y_m$ will be a local minimum).
This last event occurs simply with probability $1/2$. Hence, the probability that the number of minima $N_{\min}$ in the original walk
till its first-passage time exceeds $m$ is simply 
\begin{equation}
{\rm Prob.}(N_{\min}\ge m)= \sum_{k=m}^{\infty} Q^{({\rm fp})}(0,k)= \frac{1}{2}\, q_m \, ,
\label{cum_min.1}
\end{equation}
where $q_m$ is the probability that the auxiliary walk stays non-negative up to step $m$ and the factor $1/2$
comes from the fact that the $m$-th position of the auxiliary walk must be a local minimum of the original walk.
However, the celebrated Sparre Andersen theorem \cite{SA} tells us that 
the survival probability $q_m$ of the auxiliary walk in Eq. (\ref{y_lange.1}) up to step $m$, starting at the origin, is universal, i.e., independent
of the jump distribution $\psi(\xi)$ as long as it is symmetric and continuous. Indeed we have proved above that $\psi(\xi)$ is symmetric, continuous 
and normalized to unity. Hence we can apply the Sparre Andersen theorem to this auxiliary walk. The Sparre Andersen result says that the
survival probability $q_m$ up to step $m$  for a random walk starting at the origin is given by~\cite{SA}
\begin{equation}
q_m = {2m \choose m}\, 2^{-2\, m}\, \quad m=0,1,2,\ldots \, .
\label{SA.1}
\end{equation}
Hence, using Eq.~(\ref{cum_min.1}), the probability of having exactly $m$ minima up to the first-passage time, for $m\ge 1$,  
is given~by
\begin{equation}
Q^{({\rm fp})}(0,m) = \frac{1}{2}\, [q_m-q_{m+1}]= \frac{1}{2^{2m+2}}\, \frac{(2m)!}{m!\, (m+1)!}\, 
\quad {\rm for} \quad m\ge 1\, ,
\label{prob_nmin.1}
\end{equation}
where we used the expression of $q_m$ in Eq. (\ref{SA.1}). Note that one has to be a bit careful for the special case $m=0$, i.e., configurations
where there is no minimum. Indeed, extending this result to $m=0$, it predicts that $Q^{({\rm fp})}(0,0) = 1/4$. However, this does not include the case
where the walker jumps to the negative side at the first step, which happens with probability $1/2$. Hence, adding this contribution, we get
\bea \label{Qfp_m0}
Q^{({\rm fp})}(0,m=0) = \frac{3}{4}
\eea
Using these results in Eqs. (\ref{prob_nmin.1}) and (\ref{Qfp_m0}), one can check that $Q^{({\rm fp})}(0,m)$ is normalized to unity, i.e. 
\bea \label{norm_to_unity}
\sum_{m=0}^\infty Q^{({\rm fp})}(0,m) =1 \;. 
\eea
This then proves the universal result announced in Eq. (4) of the main text.

\begin{figure}[t]
\includegraphics[width=0.5\linewidth]{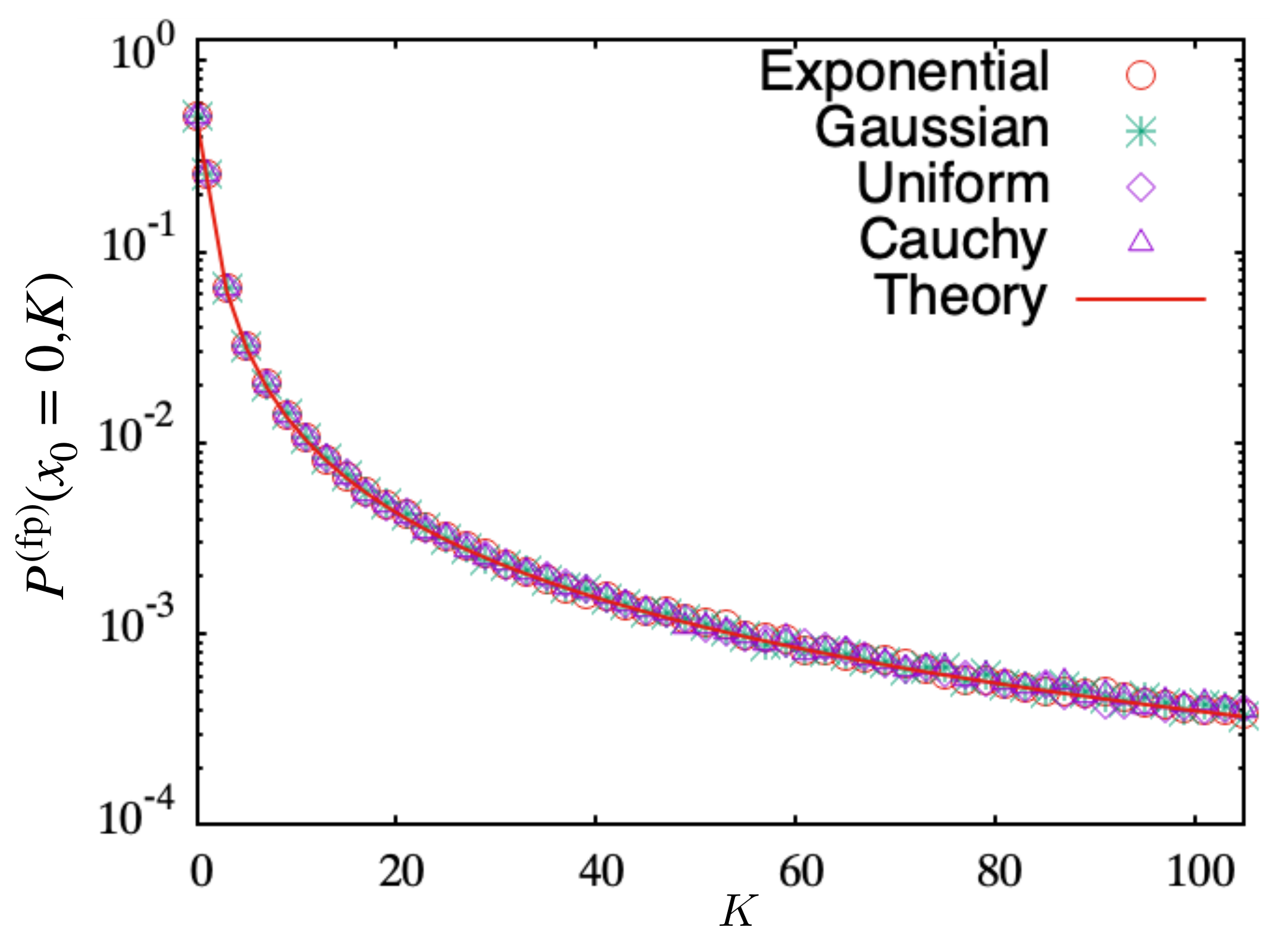}
\caption{Plot of the numerically obtained $P^{({\rm fp})}(x_0=0,K)$ for four different jump distributions, compared to the theoretical expression in Eqs.~\eqref{P_nstat.3} and \eqref{P_nstat.4}, plotted with a solid line, showing a perfect agreement. The collapse of the data for the different jump distributions on a single curve
clearly demonstrates  the universality of $P^{({\rm fp})}(0,K)$.}\label{FigPfpkN}
\end{figure}
\vspace*{0.5cm}
\noindent{\bf Universality of the distribution of the number of stationary points $P^{({\rm fp})}(0,K)$.} The argument used above for the number of minima can be extended also to prove the universality of the number of stationary points $K$ till the first-passage time. The argument proceeds as follows. It is convenient again to examine a typical trajectory as in Fig. \ref{Fig_rw_block}. This figure provides a configuration where there are at least $m$ number of minima. Next we observe that every local minimum in this configuration is preceded by a local maximum. Thus this configuration has exactly $m$ maxima. In other words, the number of stationary points $N_{\rm stat}$ in this configuration is at least $2m$. It then follows from Eq. (\ref{cum_min.1}) that 
\bea \label{P_nstat}
{\rm Prob.}(N_{\rm stat} \geq 2m) = \frac{1}{2} q_m \;,
\eea
where $q_m$ is given in Eq. (\ref{SA.1}). Setting $m = \ell -1$ with $\ell \geq 1$, we get
\bea \label{P_nstat.2}
{\rm Prob.}(N_{\rm stat} \geq 2\ell -2) = \frac{1}{2} q_{\ell - 1} \;.
\eea
We now recall that the number of stationary points till the first-passage time is always an odd number. Hence Eq.~(\ref{P_nstat.2}) shows
that
\bea \label{P_nstat.3}
{\rm Prob.}(N_{\rm stat} = 2\ell -1) = P^{({\rm fp})}(0,K=2\ell -1) = \frac{1}{2} \left(q_{\ell-1} - q_{\ell} \right) \quad, \quad {\rm for} \quad \ell \geq 1 \;. 
\eea 
Note that the result for $K=0$ is different. The case $K=0$, i.e., no stationary point till the first-passage time can happen only if the walk jumps
to the negative side after the first step itself. Any other configuration will have $K>0$. Since the probability that the walker crosses to the negative side 
after the first step is simply $1/2$, we get 
 \bea \label{P_nstat.4}
{\rm Prob.}(N_{\rm stat}=0) =  P^{({\rm fp})}(0,K=0) = \frac{1}{2} \;. 
\eea
Note that these results for $P^{({\rm fp})}(0,K)$ in Eqs. (\ref{P_nstat.4}) and  (\ref{P_nstat.3}) match perfectly with the more direct exact results obtained from
the solution of the integral equation for the special double-exponential jump distribution in Eqs. (\ref{sol:P(0,m)-fr0}) and (\ref{sol:P(0,m)-fr}). Finally, in Fig. \ref{FigPfpkN}, we compare these theoretical predictions in Eqs. (\ref{P_nstat.3}) and  (\ref{P_nstat.4}) with numerical simulations for four different jump distributions, showing a perfect agreement.

As a final remark, we note that these universal results for random walk landscapes in the first-passage ensemble can be directly transported to the RTP problem. Consider an RTP in $d$ dimensions till its $x$-component crosses the origin for the first time. In fact, the result for $P^{{\rm fp}}(0,K)$ in Eq. (\ref{P_nstat.3}) and $Q^{{\rm fp}}(0,m)$ in Eq. (\ref{prob_nmin.1}) directly hold for this RTP problem. This is unlike the fixed $t$ ensemble of RTP, discussed in subsection~\ref{sec:RTP}, where we had to average over the distribution of the number $N$ of runs. In the first-passage ensemble, since the crossing-time is summed over, we do not need any additional averaging and these results thus hold directly for the $x$-component of the RTP till its first-passage time to the origin. Hence, for the RTP, these results are also universal, i.e., independent of the spatial dimension $d$ and the velocity distribution $W(|{\bf v}|)$.